\documentclass[twocolumn]{aastex631}
\usepackage[yyyymmdd,hhmmss]{datetime}
\usepackage{CJK}
\usepackage{float}
\usepackage{booktabs} %For commands used in the tables
\usepackage{xargs} % Use more than one optional parameter in a new commands
\usepackage{chemformula}
\usepackage{graphicx}
\usepackage{xcolor, soul}
\usepackage[colorinlistoftodos,prependcaption,textsize=tiny,textwidth=.5in]{todonotes}
\usepackage{xspace}

\newcommand{\sequencer}{\texttt{Sequencer}\xspace}
\begin{document}

\title{Sequencing Silicates in the IRS Debris Disk Catalog I: Methodology for Unsupervised Clustering}

\author[0000-0001-9352-0248]{Cicero X. Lu}
\altaffiliation{Gemini Science Fellow}
\affiliation{Gemini Observatory/NSF NOIRLab, 670 N. A’ohoku Place, Hilo, Hawai’i, 96720, USA}
\affiliation{Department of Physics and Astronomy, The Johns Hopkins University, 3400 N.~Charles Street, Baltimore, MD 21218, USA}
\email{cicero.lu@noirlab.edu}

\author[0000-0002-8026-0018]{Tushar Mittal}
\affiliation{Department of Geosciences, Pennsylvania State University, 309 Deike Building, State College, PA 16801, USA}

\author[0000-0002-8382-0447]{Christine H. Chen}
\affiliation{Space Telescope Science Institute, 3700 San Martin Dr., Baltimore, MD 21218, USA}
\affiliation{Department of Physics and Astronomy, The Johns Hopkins University, 3400 N.~Charles Street, Baltimore, MD 21218, USA}

\author[0009-0001-7058-8538]{Alexis Y. Li}
\affiliation{Department of Physics and Astronomy, The Johns Hopkins University, 3400 N.~Charles Street, Baltimore, MD 21218, USA}

\author[0000-0002-5885-5779]{Kadin Worthen}
\affiliation{Department of Physics and Astronomy, The Johns Hopkins University, 3400 N.~Charles Street, Baltimore, MD 21218, USA}

\author[0000-0001-9855-8261]{B. A. Sargent}
\affiliation{Space Telescope Science Institute, 3700 San Martin Dr., Baltimore, MD 21218, USA}
\affiliation{Department of Physics and Astronomy, The Johns Hopkins University, 3400 N.~Charles Street, Baltimore, MD 21218, USA}

\author[0000-0002-9548-1526]{Carey M. Lisse}
\affiliation{Johns Hopkins University Applied Physics Laboratory, 11100 Johns Hopkins Rd, Laurel, MD 20723, USA}

\author[0000-0003-4520-1044]{G. C. Sloan}
\affiliation{Space Telescope Science Institute, 3700 San Martin Dr., Baltimore, MD 21218, USA}
\affiliation{Department of Physics and Astronomy, University of North Carolina, Chapel Hill, NC 27599-3255, USA}

\author[0000-0003-4653-6161]{Dean C. Hines}
\affiliation{Space Telescope Science Institute, 3700 San Martin Dr., Baltimore, MD 21218, USA}

\author[0000-0001-8302-0530]{Dan M. Watson}
\affiliation{Department of Physics and Astronomy, University of Rochester, 500 Wilson Blvd, Rochester, NY 14627, USA}

\author[0000-0002-4388-6417]{Isabel Rebollido}
\affiliation{ESA Research Fellow, ESAC}
\affiliation{European Space Astronomy Centre (ESAC), Camino bajo del Castillo, s/n Urbanización Villafranca del Castillo, Villanueva de la Cañada, E-28692 Madrid, Spain}

\author[0000-0003-1698-9696]{Bin B. Ren}
\altaffiliation{Marie Sk\l odowska-Curie Fellow}
\affiliation{Universit\'{e} C\^{o}te d'Azur, Observatoire de la C\^{o}te d'Azur, CNRS, Laboratoire Lagrange, Bd de l'Observatoire, CS 34229, 06304 Nice cedex 4, France}
\affiliation{Max-Planck-Institut f\"ur Astronomie (MPIA), K\"onigstuhl 17, D-69117 Heidelberg, Germany}

\author[0000-0003-1665-5709]{Joel D. Green}
\affiliation{Space Telescope Science Institute, 3700 San Martin Dr., Baltimore, MD 21218, USA}

\shorttitle{}
\shortauthors{Lu et al.}
\correspondingauthor{Cicero X. Lu}

\begin{abstract}
Debris disks, which consist of dust, planetesimals, planets, and gas, offer a unique window into the mineralogical composition of their parent bodies, especially during the critical phase of terrestrial planet formation spanning 10 to a few hundred million years. Observations from the \textit{Spitzer} Space Telescope have unveiled thousands of debris disks, yet systematic studies remain scarce, let alone those with unsupervised clustering techniques.
This study introduces \texttt{CLUES} (CLustering UnsupErvised with Sequencer), a novel, non-parametric, fully-interpretable machine-learning spectral analysis tool designed to analyze and classify the spectral data of debris disks. \texttt{CLUES} combines multiple unsupervised clustering methods with multi-scale distance measures to discern new groupings and trends, offering insights into compositional diversity and geophysical processes within these disks. 
Our analysis allows us to explore a vast parameter space in debris disk mineralogy and also offers broader applications in fields such as protoplanetary disks and solar system objects. This paper details the methodology, implementation, and initial results of \texttt{CLUES}, setting the stage for more detailed follow-up studies focusing on debris disk mineralogy and demographics.
\end{abstract}
\turnoffedittwo
\section{Introduction}
Debris disks are planetary systems that contain dust, planetesimals, planets, and gas \citep{Hughes+18}, and can have lifetime of 5 Myr to 5 Gyr \citep[][]{Hernandez+07, Wyatt+08, Chen+20}. As debris disks' life span overlaps with a critical stage of forming terrestrial planets \citep[10--200 Myr, ][]{Chambers13, Genda+15, Quintana+16}, planetary systems' formation and evolutionary history are imprinted in these disks. Amongst all components in a debris disk, dust grains reveal particularly unique information about their parent planetesimals mineralogical composition. We can directly observe the composition of dust grains formed from planetesimal collisions, in the same planetary system, and thus infer what minerals that the planets are made of. This information provides us critical insights about the properties of starting material (e.g., Fe and water content, redox properties; \citealp{hirschmann2022magma,young2023earth}). Additionally, the process of planetary differentiation (e.g., partial melting, magma crystallization, volcanic eruptions, impact associated vaporization \& melting), surface processes (e.g., weathering, aqueous alteration), and tectonics (e.g., metamorphism) produce a wide diversity of silicate mineral assemblages in the solar system \citep{Best13,hazen2023evolution}. These silicate rocks can be distinct from the bulk composition of Earth’s mantle and/or undifferentiated planetesimals \citep{Kleine+20}. Thus, the presence of different minerals in debris disks can be directly related to analogous geophysical processes on terrestrial planets, Jovian and Saturnian moons, asteroids, and comets \citep{Morlok+14, Lisse+07Comp}. 

\begin{figure}[ht!]
    \epsscale{1.0}
\plotone{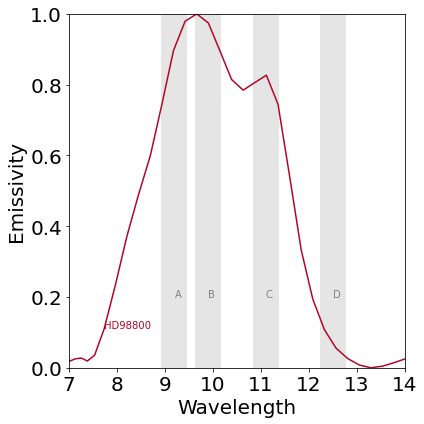}
    \caption{Spectral Indices Band Locations with an example Spitzer IRS spectrum. The band positions of the strongest $10\mu$m band is plotted against the emissivity of band A ($8.9\,$--$\,9.6\,\mu$m), B ($9.8\,$--$\,10.2\,\mu$m), C ($10.8\,$--$\,11.4$$\,\mu$m) and D ($12.2\,$--$\,12.7$$\,\mu$m). The x-axis is wavelength in microns and y-axis is emissivity which is usually defined to be disk flux divided by fitted continuum flux from $8$--$13\,\mu$m \citep{Morlok+14}. }
    \label{fig:indicesm}
\end{figure}

However, to date, there have been few systematic studies on debris disk mineralogy and structure, leveraging nearly a thousand debris disks found with the \textit{Spitzer} Space Telescope \citep[e.g.,][]{Chen+14, Mittal+15,Chen+20}, especially analyzing this dataset with modern unsupervised clustering techniques to analyze the disk population characteristics \citep[e.g.,][]{nielsen2016hierarchical, Baron+21}. Detailed spectral modeling of a small number of disks has revealed the existence of two broad categories of dust compositions - crustal-like and mantle-like compositions \citep[e.g.,][]{ Lisse+12, Morlok+14, Lisse+20weathering,Wilson+16}. The first ``crustal-like'' group is representative of terrestrial crustal materials and displays strong $9\,$--$\,9.5\,\mu$m features produced from silica, such as tektite, SiO$_{2}$, and obsidian \citep{Lisse+07Comp, Lisse+20weathering}. High-temperature processing of silica-rich differentiated crustal rocks such as evaporation and condensation during grazing hit-and-run collisions can produce these glassy minerals. The other ``mantle-like'' group resembles Earth mantle materials and contains silicates, which have strong pyroxene, and olivine bands in 9 to 12 $\mu m$ regions \citep[e.g.,][]{Chen+07, Olofsson+12, Lu+22}. These minerals can either be excavated from deep planetary interiors in giant collision events, such as the collision between proto-Earth and a Mars-sized planetesimal that formed the moon \citep{Chambers+98, Kenyon+Bromley06,gabriel2023role} or be representative of dust from undifferentiated planetesimals \citep{weiss2013differentiated,first2023galaxy}. \citet{Morlok+14}'s work reveals two groups of disks that represent analogs of crustal and mantle composition of terrestrial planets, laying the foundation for a global, process-based, classification of debris disks. However, their analysis \citep[and similar work by ][]{Lisse+12} have focused on using spectral index based approaches. In Figure~\ref{fig:indicesm}, we show the spectral indices band locations in gray, using an example spectrum from HD 98800, a young disk \citep[$\sim$10 Myr old;][]{zuniga2021hd}. The positions of the strongest $10\mu$m band are determined and integrated line fluxes are computed for bands A, B, C, and D, which are centered at $9.0$--$9.6$, $9.8$--$10.2\,\mu$m, $10.8$--$11.4$ and $12.2\,$--$\,12.7$$\,\mu$m, respectively. Subsequently, band ratio (A-B)/(C-D) along with the position of the strongest spectral feature can be calculated to distinguish composition between the crustal and mantle-like compositions on a 2D plot.

However, it is challenging to directly apply the empirical indices-based method in \citet{Morlok+14} (and similar studies) to a larger sample that spans over a wide wavelength range. There are, in particular, 2 challenges associated with the indices method: (1) incomplete understanding of the line-to-continuum ratio in the 10 $\mu$m region \& its effect on the indices for weaker features and (2) limited wavelength range used in the spectral indices (i.e., not using the full IRS $5\,$--$\,35\,\micron$ spectra). With regards to the first challenge -- the fluxes of band ratio and positions of the strongest feature are sensitive to the assumption of the underlying blackbody emission. The sensitivity of the strongest band wavelength position with respect to a $\sim100\,$K offset in temperature can be as large as $0.2\,\mu$m \citep{Chihara+02, Koike+03, deVries+12, Lu+22}.
The indices method is also limited to the $10~\mu$m region with a $\delta \lambda = 5~\mu$m (or FWHM\,=$\,2.5\,\mu$m). However, the $10\mu$m feature is empirically known to be associated with $20$ and $30\mu$m complexes \citep{Chihara+02, Koike+03, Zeidler+15}, and the $10$, $20$ and $30\,\mu$m jointly reveal information such as Fe/Mg ratio and grain temperature distribution that are inaccessible to using $10\,\mu$m alone. Because the \textit{Spitzer} IRS data contains a full wavelength range from $5$ to $35\,\mu$m, we need methods that can utilize this much wider range wavelength to extract more information about the dust composition and understand the geophysical processes associated with their formation.

With the accelerated discoveries of debris disks over the past decade, increase in available laboratory measurements, and a self-consistent and homogenous dataset for all disk spectra, we can analyze the population statistics of disk compositional properties and planetary compositional variations. However, given that we are working with disk samples with 100s of disks, we need automated methods to analyze the spectra. We would also ideally like to reduce any pre-existing human bias in disk data classification/clustering. These challenges motivate the new tool development described in this study. We develop a new non-parametric and fully-interpretable machine-learning spectral analysis tool - \texttt{CLUES} (CLustering UnsupErvised with Sequencer).
%% Let's stick with \texttt{CLUES} instead of SEQUENCES (Sequencer combined with  Unsupervised clustering -- SEQUEnciNg ClustEring unSupervised
This allows us to discover new groupings and trends in spectra data by combining multiple unsupervised clustering methods with multi-scale distance measures \citep[\sequencer algorithm,\,][]{Baron+21}. 

While the primary focus of this study is Spitzer IRS data and debris disk science, our toolkit can be applied to other high-dimensional spectral datasets. Thus, the method has relevance for not only other fields within planet formation such as protoplanetary disk spectra and exoplanets emission/transmission spectra but also to other mineral spectroscopy in more general areas of astrophysics and remote sensing (e.g., high-resolution satellite-based imaging). For instance, this tool suite would be useful for multi-spectral astrophysical datasets such as high-resolution optical spectroscopy (e.g., Keck and LSST - stellar data). Similarly, \textit{JWST} MIRI instrument operates in commensurate wavelengths as that of the \textit{Spitzer} IRS. Thus, our tool can be directly applied to \textit{JWST} MIRI data without significant additional customization. Finally, multi-spectral and hyperspectral datasets are increasingly becoming common in terrestrial and planetary remote sensing (e.g., visible and near-IR hyperspectral data, PACE satellite). 

A key utility of this tool is to reduce the effective dimensionality of the datasets, especially when they contain 100s to 10,000s of individual spectra (e.g., Integral field spectrograph for \textit{JWST} or data hypercubes for hyperspectral mapping of planetary surfaces) and find representative end-member spectra that can be analyzed in detail with detailed mineralogical modeling and follow-up observations. In the context of debris disks, we are interested in finding representative exemplar disks for detailed follow-up observations with \textit{Hubble}, \textit{JWST}, ALMA, and other optical-IR observatories - such that we can learn about the planet formation processes in general rather than being just focused on interesting, albeit outlier, systems. 

In this work, we describe the methodology for the unsupervised clustering. In Section~\ref{section:data}, we briefly outline our overarching spectral analysis architecture which consists of four main parts. We describe the part 1 (P1), the data preprocessing stage in Section~\ref{section:preprocessing}. We then describe P2, the \texttt{CLUES} workflow in detail in Section~\ref{section:analysis} and show clustering results of applying \texttt{CLUES} to 3 different datasets in Section~\ref{section:results} - a pure material library, an ensemble of 59 meteorite spectra, and one debris disk spectrum along with the Emissivity library. In Section~\ref{section:discussion}, we discuss the implications and potential applications for \texttt{CLUES}. Finally, we summarize the paper in Section~\ref{section:conclusion}. Our results establish the analysis framework to analyze debris disk mineralogy in Paper II.  

\section{Spectral analysis workflow}\label{section:data}
In this section, we summarize the overall workflow as an overview with the next sections laying out the parts: 
\begin{itemize}
    \item P1. Preprocessing stage (getting from observations to emissivity) e.g., Stellar Photosphere Modeling and Subtraction, Average Emissivity calculations, Disk Continuum Modeling - different approaches for continuum subtraction followed by data binning and normalization. 
    \item P2. \texttt{CLUES} Data analysis Workflow for unsupervised clustering analysis
    \item P3. Tools and methods for Visualization of results as well as various intermediate steps in the workflow to better interpret the results.
    \end{itemize}
The techniques described here in parts P1 and P2 will be tailored for IRS/mid-IR data mostly and will need some modifications for other datasets depending on the science applications.
% Here, it would be useful to briefly mention that Stage a1, a2 would be dependent on the specific dataset of interest and would need to be modified for different applications. 

% Figure~2 needs to be updated. 
\begin{figure}[ht!]
    \centering
\includegraphics[width=\linewidth]{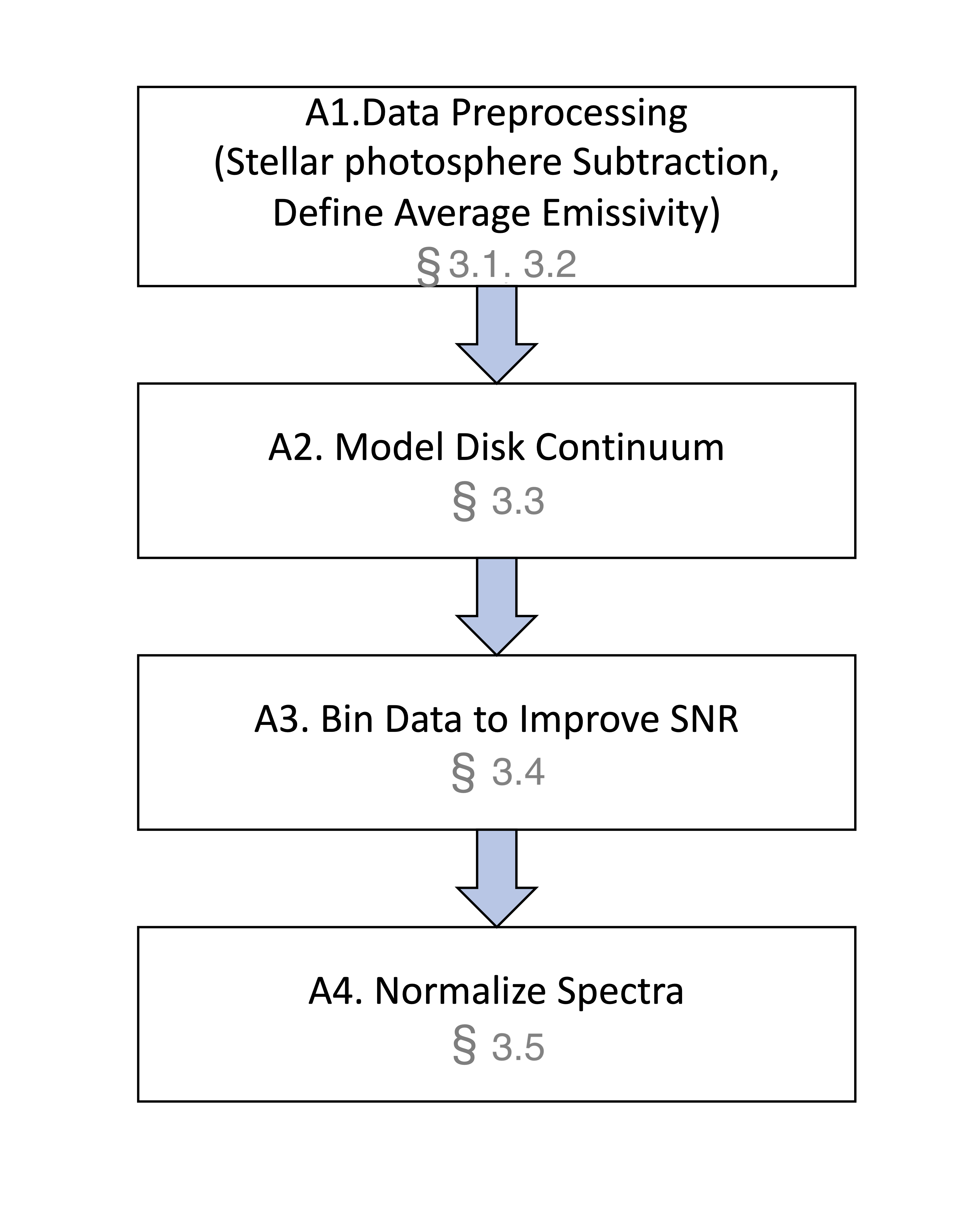}
\caption{A Flowchart of Data Processing Steps (P1): Each rectangular box represents a data processing step (in black fonts) and its corresponding subsections (in gray fonts) in the next section.}\label{fig:data-flowchart}
\end{figure}

\begin{figure}[th!]
    \epsscale{1.2}
    \plotone{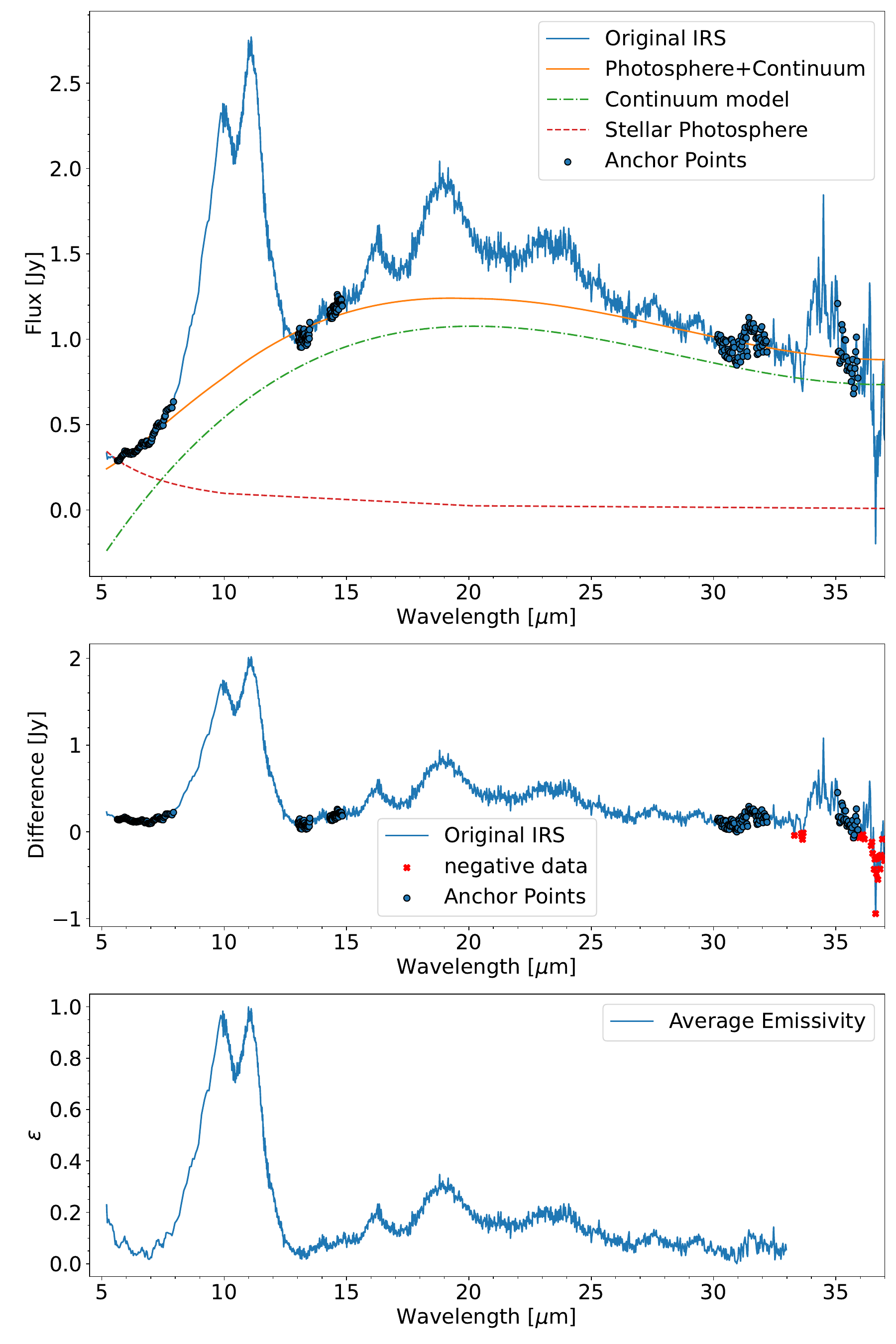}
    \caption{Visualization of Data Processing (Section~\ref{section:preprocessing}) for an example debris disk spectra, HD 113766. \textbf{Top}: The original spectrum of HD 113766. We show the stellar photosphere (red dotted line), disk continuum (green dashed line), anchor points (black points) for fitting the disk continuum, and the sum of the stellar and disk flux contribution (orange solid line) as described in sections \ref{StellarContinuum} and \ref{sub:diskContinuum}. \textbf{Middle}: We present a spectrum in blue, with the stellar photosphere (red dotted line) and the debris disk continuum (green dashed line) subtracted. The red points denote where data points become negative in value after disk continuum flux subtraction as described in \ref{StellarContinuum}. We truncate the spectrum at wavelengths beyond 33 $\mu$m where the photosphere-subtraction and disk-continuum subtraction often result in negative data due to noises in \textit{Spitzer/IRS} data. \textbf{Bottom}: We show an emissivity spectrum in blue as described in Section~\ref{sub:emissivity}. 
    }
    \label{fig:continuum}
\end{figure}
\section{Preprocessing Stage 1}\label{section:preprocessing}
 We then describe our photosphere fitting procedure to isolate the disk emission from the stellar emission. Next, we define the concept of ``average emissivity'' for debris disk spectra. Thereafter, we describe our data pre-processing steps, including normalization and binning, to improve the SNR of the data.  The entire process is illustrated in the flowchart in Figure~\ref{fig:data-flowchart} and accompanied by step-by-step illustrations with a debris disk example in Figure~ \ref{fig:continuum}. 

\subsection{Stellar Photosphere Modeling and Subtraction}\label{StellarContinuum}
The debris disks in the \textit{Spitzer} IRS catalog are spatially unresolved and therefore the spectra contain flux both of stellar photospheres and of the disk thermal emission, such that 
\begin{equation}
    F_{target}(\lambda) = F_{\star}(\lambda) + F_{disk}(\lambda)
\end{equation}
To obtain a spectrum with only disk emission, we subtract the stellar photosphere emission from the disk spectra. The photosphere parameters are taken from \citet{Chen+14}, in which they model the stellar parameters with absolute V-band magnitudes from Stromgren photometry corrected for stellar rotation. They also apply bolometric correction \citep{Flower+1996} in their luminosity derivation. There have been no new V band measurements for our sample since the publication of \citet{Chen+14}; hence we will use their best fit stellar parameters as is. 

\subsection{Determining the Average Emissivity for the Small Dust Grain Population from IRS Spectra}\label{sub:emissivity}
Emissivity is an intrinsic radiative property of dust grains that strongly depends on dust composition, shape and sizes. The flux of a debris disks is correlated with its dust grain emissivity, shape and size in the following way: 
\begin{equation}
    F_{disk, \nu} = \int_{a_{min}}^{a_{max}} \frac{\pi a^2}{d^2} \sum\limits_i^k Q_{em,i, a} \frac{dn}{da} B_{\nu}(T_i(a))da, 
    \label{eqn-disk1}
\end{equation}
where $a_{min}$ ($a_{max}$) is the minimum (maximum) size of the grain, and \textit{dn/da} is an arbitrary size distribution of grains with radius $a$. $Q_{em,i}$ is the emissivity (hence also the absorption efficiency) of a grain with $a$ given composition ($i^{th}$ composition with up to a total $k$ dust species) and size (radius), $a$. $B_{\nu}(T_i)$ is the blackbody emission for the $i^{th}$ composition. $\pi a^2$ is the geometric surface area of the grains that are emitting, assuming the grain is a simple sphere and \textit{d} is our distance to the system.

We can extract information about grain emissivity of the grains in a debris disk given its spectrum. In an ideal scenario, if we know the precise shape and size distribution, we can work out $Q_{em,i}$, the grain emissivity for each species, $i$, at a given size, a, from the smallest nanometer-sized grains to the largest km-sized planetesimals. However, realistically, the grain emission features in the MIR wavelengths are only sensitive to the composition of grains in the Rayleigh limit. In the Rayleigh limit, the grain emission features are probing grains that are much smaller than the emission, in our case, grains from sub-$\mu$m to at most a few $\mu$m in sizes. We can use IRS spectra to constrain the emissivity properties of averaged over the grain sizes and also over a range of blackbody temperatures, from $\sim 100-600$ K.

We define the average emissivity of the small (sub-$\mu$m to a few $\mu$m) grain populations in debris disks to be 
\begin{equation}
    \text{Avg Emissivity} = \Bar{\epsilon} \equiv \frac{F_{disk}}{ F_{cont}},
\end{equation}
where the $F_{cont}$ is the continuum flux that approximates the sum of the thermal emission of large grains and planetesimals' blackbody emission. This average emissivity, $\Bar{\epsilon}$, constrains the grain composition, temperature, and shape. Extensive laboratory experiments show compositions most predominately affect the locations at which the emission features peak at, while grain shape predominantly affects the overall full-width half-maximum (FWHM) of the emission features. In addition, as indicated in Equation \ref{eqn-disk1}, the grain temperature ($B_{\nu}(T_i(a)$) and overall abundance (dn/da) of grains at that temperature affect the relative amplitude of the $10$, $20$ and $30\,\mu$m emission complexes. Since the emissivities of dust grains with various compositions and sizes are measured in the lab, we can compare the average emissivity of debris disks to the lab-measured emissivities to study the dominant grain compositions and sizes in debris disks and understand their demographics. 

Since disk continuum emission is an important part of $\Bar{\epsilon}$, in the following section, we describe our disk continuum emission modeling in detail.  

\subsection{Disk Continuum Modeling}\label{sub:diskContinuum}

The objective of continuum modeling is to isolate the characteristic dust features emitted by sub-$\mu$m--sized grains from the blackbody emission of large (10--100 $\mu$m) dust grains. The large grain emission can be approximated by polynomials fitted to spectral regions that are relatively free of narrow emission features \citep{Mittal+15, Watson+09}. In Figure~\ref{fig:continuum}, we demonstrate the continuum fitting process with an example IRS disk spectra, HD 113766 from our sample. We use $5.61$--$7.94$, $13.02$--$13.50$, $14.32$--$14.83$, $30.16$--$32.19$, and $35.07$--$35.92 ~\mu$m regions (shown in blue dots in Figure~\ref{fig:continuum}) as anchoring points to fit for a 3rd order polynomial. The resulting polynomial fit is shown in the upper panel in orange and represents the contribution of large grains that emit like black bodies. It is worth mentioning that the spectra become noisy (due to instrumental and data reduction artifacts) in regions beyond $30\,\mu$m and therefore, might appear to have sawtooth patterns and negative values in flux. In the middle panel of Figure~\ref{fig:continuum}, we show the continuum-subtracted spectra, pointing to a common situation where the continuum is over-fitted partially due to the contribution of these noisy regions. As highlighted by red dots, regions beyond the $30\,\mu$m continuum model often overshoot the IRS flux. Therefore, we apply an offset value such that the offset continuum flux is always equal to or less than the IRS spectrum. We also exclude the data in regions beyond $33\,\mu$m where noises dominate the spectrum. We then divide the offset disk continuum emission (green dashed dotted line in the top panel of Figure~\ref{fig:continuum}) from the spectra to obtain emissivity spectra in the bottom panel of Figure~\ref{fig:continuum}. If throughout the spectrum wavelength range (from $7\,\mu$m out to $33\,\mu$m, beyond which the fringing takes over), there exists at least one point at which continuum subtraction results in a negative value, we take the absolute values of the resulting negative values. We use the largest absolute value as the offset value to apply to the continuum model such that there are no negative values in the spectrum after continuum subtraction.

Another choice for fitting the continuum is to use multiple black body components, because black body components are proxies to the exo-asteroidal belt and exo-Kuiper belt analogs. We also experiment with the blackbody approach, but find more than half of the IRS debris disk sample is best modeled with a continuous disk model instead of a few black bodies \citep[consistent with ][]{Mittal+15}. In addition, the emission from large dust grains population cannot be well-modeled by black body emissions. 

For the complete debris disk catalog sample which we will use in subsequent analysis, we pre-selected debris disks based on solid-state features identified by \citet{Mittal+15}. 
They define these features' signal-to-noise ratio (SNR) as the ratio of the integrated flux over the 10 and 20$\,\mu$m bands to their respective uncertainties. This method identified 120 disks with solid-state features, amenable for the disk continuum modeling methods described in the above two paragraphs.

\subsection{Binning data in Spectra to improve SNR} \label{sub:bin}
We bin the data to improve the SNR of each spectrum because at least 20\% of spectra in our sample show fringing patterns as a result of detector artifacts. We combine N adjacent data points by averaging both their wavelength and emissivity values. The uncertainties are calculated by combining the uncertainties of each data point in quadrature. We  experiment with different values of N for binning data and find N=4 optimizes the fringe reduction and only reduces spectral resolution by a factor of 4, while doubling the SNR of each data point in the binned spectrum.  

\subsection{Spectra Normalization}\label{sub:norm}
To compare the disks with differing brightness and not have our analysis being solely dominated by these variations, we normalize the spectra in the $8\,$--$\,13\,\mu$m region. In our entire dataset, we have disks with 10-$\mu$m emission flux as high as $25$ Jy and as low as $0.1$ Jy. 
% The large range in spectral flux will introduce bias when comparing spectra with each other unless accounted for. 
Since our focus is explicitly on the dust composition, rather than the amount of dust (which dominates the absolute flux), we normalize each spectrum by selecting the highest point in each emissivity spectrum from $8$--$13\,\mu$m following \citet{Morlok+14} and normalize the highest value to unity. 

One main limitation of this methodology, as applied to the \textit{Spitzer} data in particular, is the analysis of the $20$ and $30\,\mu$m wavelength features. Even though the emissivity definition has proven to be an effective method for the $10\,\mu$m features that span from $8$--$13\,\mu$m, it is not as effective for the $20\,\mu$m spectral features. As the silicate and silica emission bands are narrow ($2\,$--$\,5\,\mu$m in equivalent width) in the $10\,\mu$m region, the emitter can be considered as $300$--$500$K blackbodies. It is thus justified to divide the disk flux by the emissivity by dividing a blackbody-like continuum flux over the narrow wavelength range. However, the silicate emission bands become broad ($\geq10\mu$m in equivalent width) over the $20$ and $30\,\mu$m wavelength region, and the continuum emission cannot be approximated by a single blackbody temperature but rather a combination of various temperatures. Our lack of knowledge about the underlying continuum blackbody temperatures limits our ability to precisely determine the relative amplitudes of features. In general, the inclusion of an increasingly colder component would bring down the emissivity in the longer wavelength range and cause systematic errors in the relative amplitude of $20$ and $30\mu$m features. Thus, overall there are a few different choices for normalization and background removal for unresolved disks (e.g., most \textit{Spitzer} IRS spectra) wherein the $10$, $20$, and $30\,\mu$m features may not even represent the same physically co-located dust population. This issue is much less of a challenge for spatially resolved disks. 
For example, one way to correctly estimate the underlying continuum is to jointly perform an image and spectra analysis to determine the overall spatial distribution of various populations of dust \citep[e.g.,][]{Ballering+16}. The measured dust spatial profile will allow us to perform dust-mass-weighted normalization. Unfortunately, because only less than $10\%$ of spectra in the \textit{Spitzer}/IRS catalog is spatially resolved, for the following analysis, we will use the simple spectral normalization scheme described in the previous paragraph for simplicity and generality. 
However, for our debris disk analysis in Paper II, we explore the effect of different normalization choices on our final results and our interpretation of the dust composition. 

%[Discuss some other potential normalization choices here and what are the physical assumptions per se for each - you don't have to show all the figures here per each normalization, but this is the place to discuss some of the issues that Dr. Christine pointed out re 10 vs 20 micron (and re dust locations). Nominally, for spatially resolved systems, this becomes less of an issue.]

After performing these data processing steps - stellar photosphere subtraction, disk continuum fitting, average emissivity calculation, binning data, and spectral normalization, we display the resulting emissivity spectra in Figure~\ref{fig:continuum} bottom panel. 

\begin{figure*}[ht!]
\epsscale{1.0}
\plotone{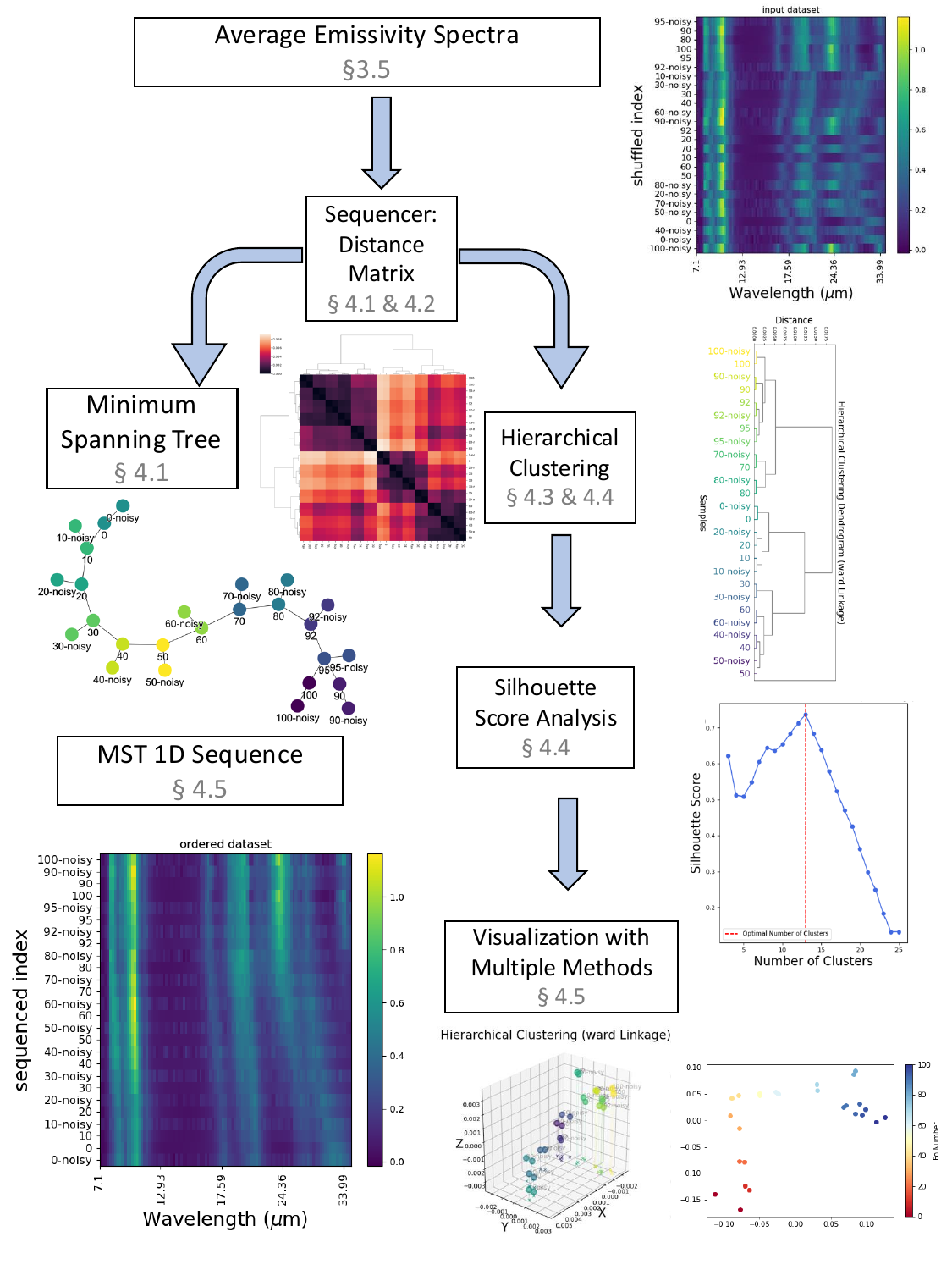}
\caption{
    \textbf{A Flowchart of Data Analyses Steps - \texttt{CLUES}}: Each rectangular box represents a data processing step (in black fonts) and its corresponding subsections (in gray fonts). The top rectangular box displays our input forsterite library emissivity spectra. The subsequent step involves calculating the distance matrices using \texttt{Sequencer}. This distance matrix enables us to perform two separate analyses, as indicated by the bifurcated arrows in the next step. We have the option to calculate a minimum spanning tree by collapsing the distance matrix into a 1D sequence. Alternatively, hierarchical clustering algorithms can be utilized to classify compositionally-representative spectra. The next box connected to the "hierarchical clustering" rectangular box corresponds to the silhouette score analysis, which serves as a clustering criterion. By applying this criterion, we can generate our final outputs, groupings of spectra for further parametric modeling. Finally, various tools are employed to visualize the distance matrices for our dataset, facilitating the understanding of the correlation between any science target spectrum and external mineral library spectra.}.\label{fig:flowchart-analysis}
\end{figure*}

\section{CLUES Analysis Workflow}\label{section:analysis}

The next step after data processing involves classification of the disks using our \texttt{Sequencer} derived workflow (Figure~\ref{fig:flowchart-analysis}). Briefly, we start by calculating a multi-scale distance matrices using \texttt{Sequencer} algorithm. This distance matrix enables us to perform two separate analyses, as indicated by the bifurcated arrows in the next step. We have the option to calculate a minimum spanning tree (MST) by collapsing the distance matrix into a 1D ordered sequence. The MST minimizes the total edge weights and connects all nodes (unique spectra in our case) without any cycles. MST thus provides one visualization of the relationship between spectra and their resemblance with respect to each other.

Alternatively, hierarchical clustering algorithms can be utilized to classify compositionally-representative spectra. The next box connected to the rectangular box corresponds to the silhouette score analysis, which serves as a cluster number selection criterion. By applying this criterion, we can generate our final outputs, which consist of groups of clustered spectra for the most likely number of groups. Finally, various tools are employed to visualize the distance matrices for our dataset, facilitating the understanding of the correlation between debris disk composition and external stellar properties. In the following, we describe each step in detail.

\subsection{Sequencer and its Associated Workflow}
% Add: Transition sentence, Need a tool that uses the full wavelength information and then can also quantify the difference between pairs of spectra and cluster them, etc

% Hence, we will use a tool, sequencer that addresses all of these aspects. 

\texttt{Sequencer} \citep{Baron+21} is a tool that provides a non-parametric and systematic way to simultaneously examine an ensemble of spectra to find underlying patterns. In this section, we run the \sequencer tool on a monomineralic spectral template library, in which each spectrum represents only one mineral composition, to compute the distance matrix, which characterizes how different or similar any pair of spectra is. 

The \sequencer method can overcome challenges posed by the indices method. While the indices method is less suitable to the $20$ and $30\,\mu$m features due to a much smaller line-to-continuum ratio and large equivalent widths, \sequencer uses information from the full wavelength range. Instead of picking out a singular strongest wavelength value across the sample, \sequencer quantifies the differences between pairs of spectra and clusters them without prior assumptions. Hence, we will use \sequencer to address these aspects. 

We demonstrate our workflow in Figure~\ref{fig:flowchart-analysis}, where each rectangular box represents a data processing step and is described in a subsection. 
The upper rectangular box illustrates our input \textit{Spitzer} IRS debris disk spectra. Following this, we calculate the distance matrices using \texttt{Sequencer}. These matrices facilitate two distinct analyses, as demonstrated by the split arrows in the subsequent step. One option is to generate a minimum spanning tree by reducing the distance matrix to a one-dimensional sequence. Alternatively, we can apply hierarchical clustering algorithms to identify spectra with representative compositions. Connected to the "hierarchical clustering" rectangle is the next box, which pertains to the silhouette score analysis, utilized as a criterion for clustering.
By applying this criterion, we can generate our final outputs, which consist of the average spectra for the clusters. Finally, various tools are employed to visualize the distance matrices for our dataset, facilitating the understanding of the correlation between debris disk composition and external stellar properties.

There are however a few caveats that are particularly important for using the \sequencer. In particular, the data needs to be appropriately formatted to use in \sequencer. \sequencer takes an array of spectra that share the exact same wavelength axis. To meet this requirement, we down-sample all spectra to the lowest resolution data in the sample, because the IRS disk spectra are observed with a combination of spectral resolutions: R$\sim100$ and R$\sim600$.  The advantage of our ``downsampling by binning'' approach is that all data points in our sample have high fidelity because we do not produce additional data points with ``interpolated'' values. 

\begin{figure*}[ht!]
\epsscale{1.1}
\plotone{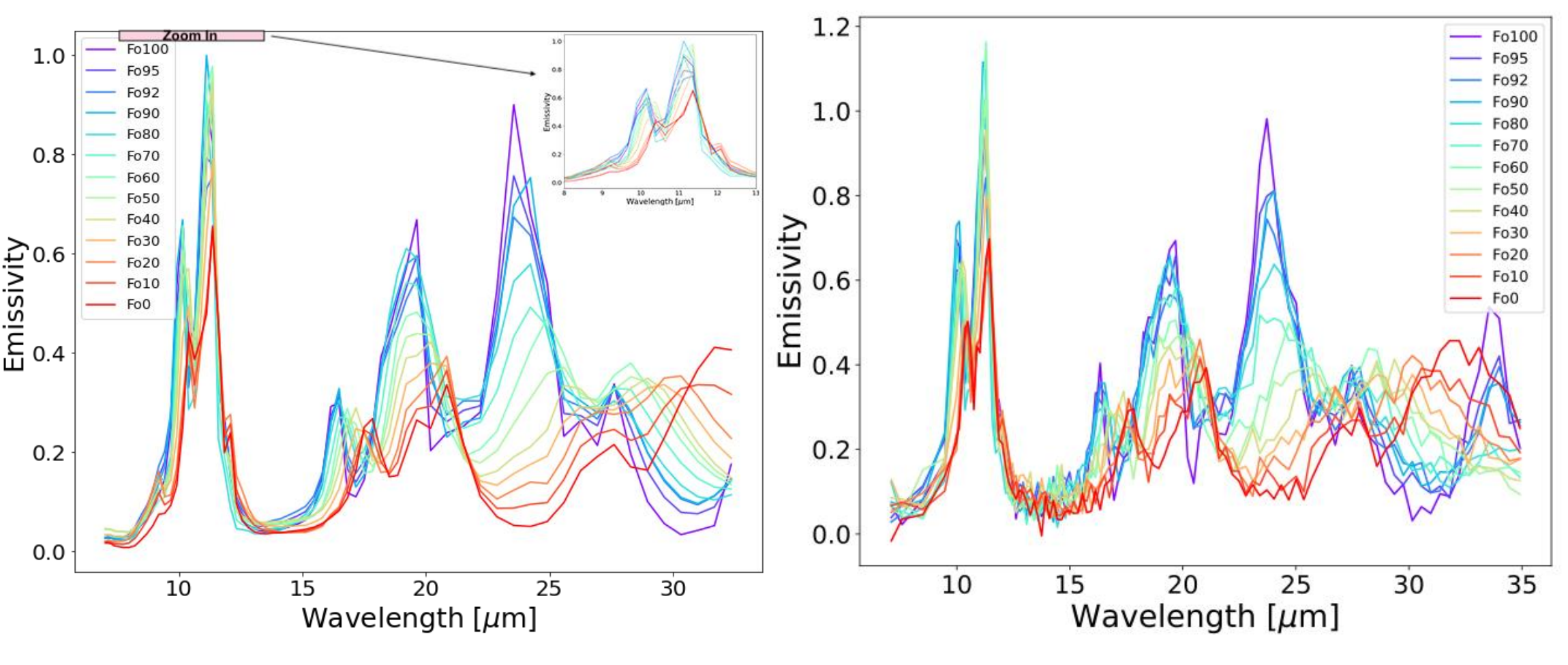}
\caption{\textbf{Forsterite Emissivity Library.} Left: Forsterite Emissivity plotted as a function of Fo number from Jena Database \citep{Chihara+02}. Right: Forsterite Emissivity with $5$\% random Gaussian noise. \label{fig:forlib}}
\end{figure*}

For the template spectral library, we use laboratory forsterite (enstatite) data \citep{Chihara+02, Koike+03} and down-sample the spectra to the IRS spectral resolution (R$\sim100$). In Figure~\ref{fig:forlib}, we show the MIR spectral library of forsterite as a function of Fo number, which characterizes  the Mg/(Mg+Fe) ratio, in forsterite grains. We also make a copy of ``noisy'' data to simulate the uncertainties in the data such as fringing. In Figure~\ref{fig:forlib} right panel, we add random Gaussian noise that is $5\%$ of the flux level as the IRS data AdOpt extraction in our sample usually corrects the effect of fringing to $<5\%$ \citep[usually close to the noise floor at $1$--$2\%$;][]{Lebouteiller+10, Higdon+04}. 

Our debris disk sample is selected from a master catalog of $571$ Spitzer IRS debris disk spectra from \citet{Chen+14}. Roughly $20\%$  ($110$ out of $571$) of the disks in the master catalog were processed with optimal extraction and the rest $80\%$ are processed with the Advanced Optimal extraction method \citep[AdOpt,][]{Lebouteiller+10} archived in the Cornell Atlas of Spitzer/Infrared Spectrograph Sources \citep{Lebouteiller+2011}. Briefly, for disks observed with optimal extraction, the sizes of the extraction window increase as a function of wavelength, consistent with the increasing size of the diffraction-limited point-spread function. For disks processed with AdOpt, in addition to the wavelength dependence on PSF, the detector-level spatial dependence on PSF is accounted for. AdOpt uses empirical super-sampled PSFs to simultaneously fit the 2D-PSF to all of the pixels on the detector in the spatial and spectral directions of the slit to account for spatial and wavelength-dependent detector biases such as inter-pixel variability and fringing. In doing so, AdOpt weights the pixels in the extraction window by their SNR and position on the detector to measure the flux at every wavelength and achieves a $\sim\,1$\,--\,$2$\% level point-to-point uncertainty on average. A detailed description of data reduction can be found in Section 4 of \citet{Chen+14}.

We select a lower and an upper limit to the wavelength range uniform across our dataset of IRS debris disk spectra, considering the spectral resolution, fringing pattern, and known issues to the data products. We select a cut-off wavelength of $7\,\mu$m at the blue end of the spectra because $5$--$7\,\mu$m region is known to have a steep declining slope for debris disks with an inconclusive cause, that could be either astrophysical or attributed to pipeline point-to-point calibration issue. We also truncate the spectra beyond $33\,\mu$m because degraded data quality longward of the wavelength. As shown in Figure \ref{fig:continuum}, the photosphere-subtraction and disk-continuum subtraction often result in negative data due to noises in Spitzer/IRS data in regions beyond $33\,\mu$m.

Although our data contains a mixture of low-resolution (R$\,\sim100$, $5.2$\,--\,$38\,\mu$m) and high-resolution (R$\,\sim600$,  $9.9$\,--\,$37.2\,\mu$m), both the modes have an overlap in the $9.9$\,--\,$37.2\mu$m wavelengths range. Additionally, the majority of spectra in our catalog were also observed in SL module such that spectral region between $7$ and $9.9$ $\mu$m is covered. In the downsampled dataset, between $7$ and $9.9$ $\mu$m, SL data was directly used and high-resolution data was used wherever applicable for longer wavelengths. Our common wavelength axis has a resolution of R$\,\sim\, 60-100$ data points, nearly equivalent to that of IRS low-resolution spectral resolution. 
% For our analysis, we limit the wavelength range to 7–33 $\mu$m, setting the lower cutoff at 7 $\mu$m due to the unexplained slope observed between 5–7 $\mu$m in several disks as well as potential challenges with fringing/high noise in spectra. The upper cutoff at 33 $\mu$m is similarly due to fringing and degraded spectral quality. Truncating beyond 33 $\mu$m prevents fringing noise from affecting the accuracy of disk continuum models. For our large IRS debris disk dataset (to be used in Paper II) the common wavelength axis has a resolution of R $\sim$ 60–100. 
However, the CLUES algorithm can be used for different spectral resolution spectra depending on the use cases.

\subsection{Selecting a Distance Metrics}\label{subsec:distance-metrics}
We use the Earth-Mover-Distance (EMD) metric which is a measure of the distance between two probability distributions over a spectral region. EMD \citep{szekely2013energy} first calculates a probability distribution function (PDF) for the entire dataset. Then for each individual spectrum, it calculates the distance between the spectrum and the PDF. A pairwise distance between two spectra is computed by taking the squared sum of their individual distances to the PDF. In our situation, this distance can be understood as the product of the FWHM of spectral features and its spectral width for every two disks in the sample. It then sorts the disks according to that distance to reveal any interesting trends across the sample. In comparison, L2 is simply the Euclidean distance often used in $\chi^2$ minimization. It's defined as the distance between two points on an Euclidean space and its mathematical expression can be found in \citet{Baron+21} Equation (6). While there are alternative distance measures (e.g., Euclidean distance or energy distance between spectra) possible, we empirically find that the EMD maximizes the minimum spanning tree elongation (implying efficient low dimensional projection while keeping the most meaningful variation within a high-dimensional dataset) among the commonly available distance measures. We jointly explore the combinations of the distance metrics with distance scale expand on more details in the next section (Sect. 
\ref{subsec:distance-scale}).

\subsection{Optimizing the Distance Scale}\label{subsec:distance-scale}
We focus on comparing the emission lines in debris disk spectra to classify the disk mineralogy. Emission line strengths are empirically quantified by their equivalent width and line-to-continuum ratio in parametric models. Because the \texttt{Sequencer} is a non-parameter algorithm, it introduces a new concept called ``distance scale'' (\textit{l}).

The distance scale \textit{l} decides the minimum wavelength range over which to compare any two spectra and therefore is an important parameter to investigate. Before computing the distance matrix, the \texttt{Sequencer} divides a spectrum of N data points into N/2\textit{l} contiguous chunks, where \textit{l} denotes the distance scale. \texttt{Sequencer} then computes the distance metric with respect to the chunks that one defines. The distance matrix focuses on the broad spectral feature with a small scale number and on narrow spectral features with a large scale number. In Table \ref{tbl-elon-stats}, we show the results after experimenting with \textit{l} of 1, 2, 5, 10, and 20 with EMD and L2 distance metrics. We find overall, EMD performs better than L2 on all scales. For EMD, we find \textit{l}= 5 empirically gives the best elongation that produces the longest sequence for the sample library spectra of forsterite. Therefore, a scale of 5 for the spectral library dataset means that the entire spectrum from $7$--$33$ $\mu$ms is divided into 30 chunks each with a wavelength range that spans  1\,$\mu$m. 

\begin{figure*}[th!]
    \epsscale{1}
    \plotone{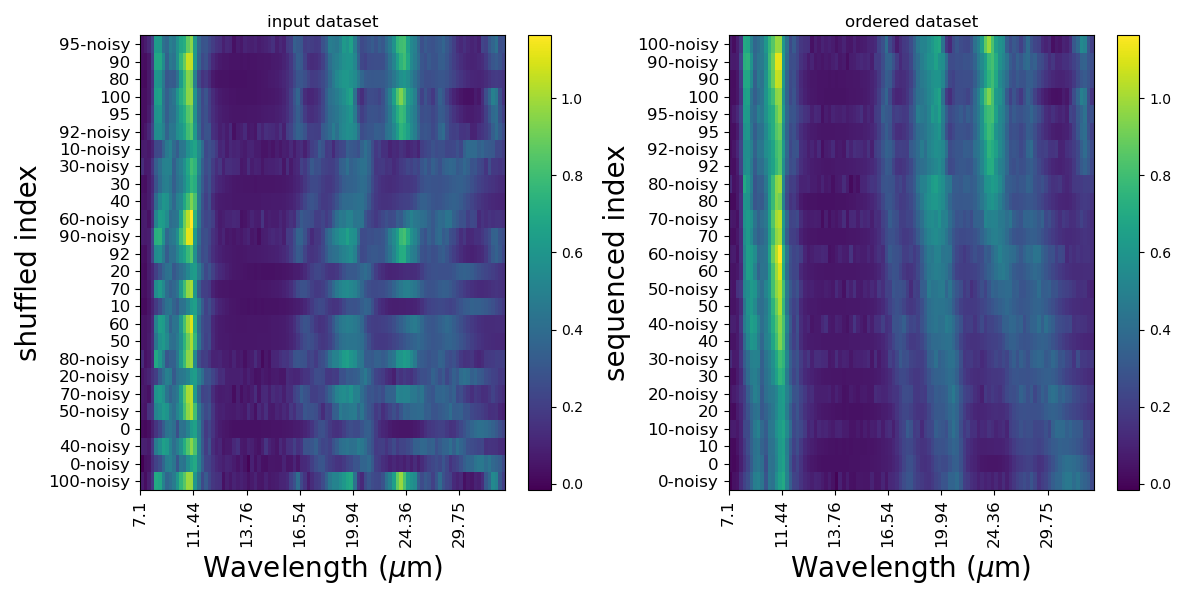}
    \caption{\textbf{Input data versus output sequenced spectra with the best elongation}. Left:Input Forsterite spectral Library that includes both panels in Figure~\ref{fig:forlib}. Right: Output sequence selected by the best elongation with an EMD metric and a scale of 5.}
    \label{fig:forlib_out}
\end{figure*}

\begin{deluxetable}{lcc}
\tablecaption{Elongation for Various Metrics and Scales}\label{tbl-elon-stats}
\tablehead{
\colhead{Metric} & \colhead{Scale (l)
} & \colhead{Elongation}}
\startdata
EMD & 1 & 15.11\\
EMD & 2 & 16.57\\
EMD & 5 & 21.49\\
EMD & 10 & 12.22\\
EMD & 15 & 8.91\\
EMD & 20 & 8.91\\
EMD & 50 & 9.57\\
L2 & 1 & 8.91 \\
L2 & 2 & 8.91\\
L2 & 5 & 8.91\\
L2 & 10 & 9.74\\
L2 & 15 & 8.91\\
L2 & 20 & 8.91\\
L2 & 50 & 9.57
\enddata
\tablecomments{EMD stands for Earth Mover Distance metric, L2 stands for the Euclidean distance and \textit{l} stands for distance scale. The details for the two metrics and distance scale are presented in sections \ref{subsec:distance-metrics} and \ref{subsec:distance-scale}. The table uses the data shown in Figure \ref{fig:forlib_out}.}
\end{deluxetable}

In Figure \ref{fig:forlib_out}, we show the collapsed 1D sequence with an maximized elongation of $21$ out of a total of $26$ forsterite library spectra. In Fig. \ref{fig:forlib_out} left panel, we show input data of ``clean'' and ``noisy'' pairs of forsterite spectra with various Fo numbers. The input spectra are shuffled with no specific orders. Fig. \ref{fig:forlib_out} right panel shows when \textit{l}=5 and EMD is selected, the elongation of the sequence across the dataset. Pairs of ``clean'' and ``noisy'' forsterite spectra are neatly ordered with an increasing Fo number as indicated on the y axis.

However, while a single scale of 5 optimizes the output 1D sequence, we use the aggregate information from a list of scales to understand both narrow and broad features. We use the weighted distance matrix calculated from a number of elongation-weighted distance matrices \citep[see Eq (4) in ][]{Baron+21}. In the next step, we visualize the weighted distance matrix over a scale list of [1, 2, 5, 10, 20, 50]. 

\subsection{Clustering analysis}\label{sec:clustering}
\begin{figure*}[ht!]
    \epsscale{1.2}
    \plotone{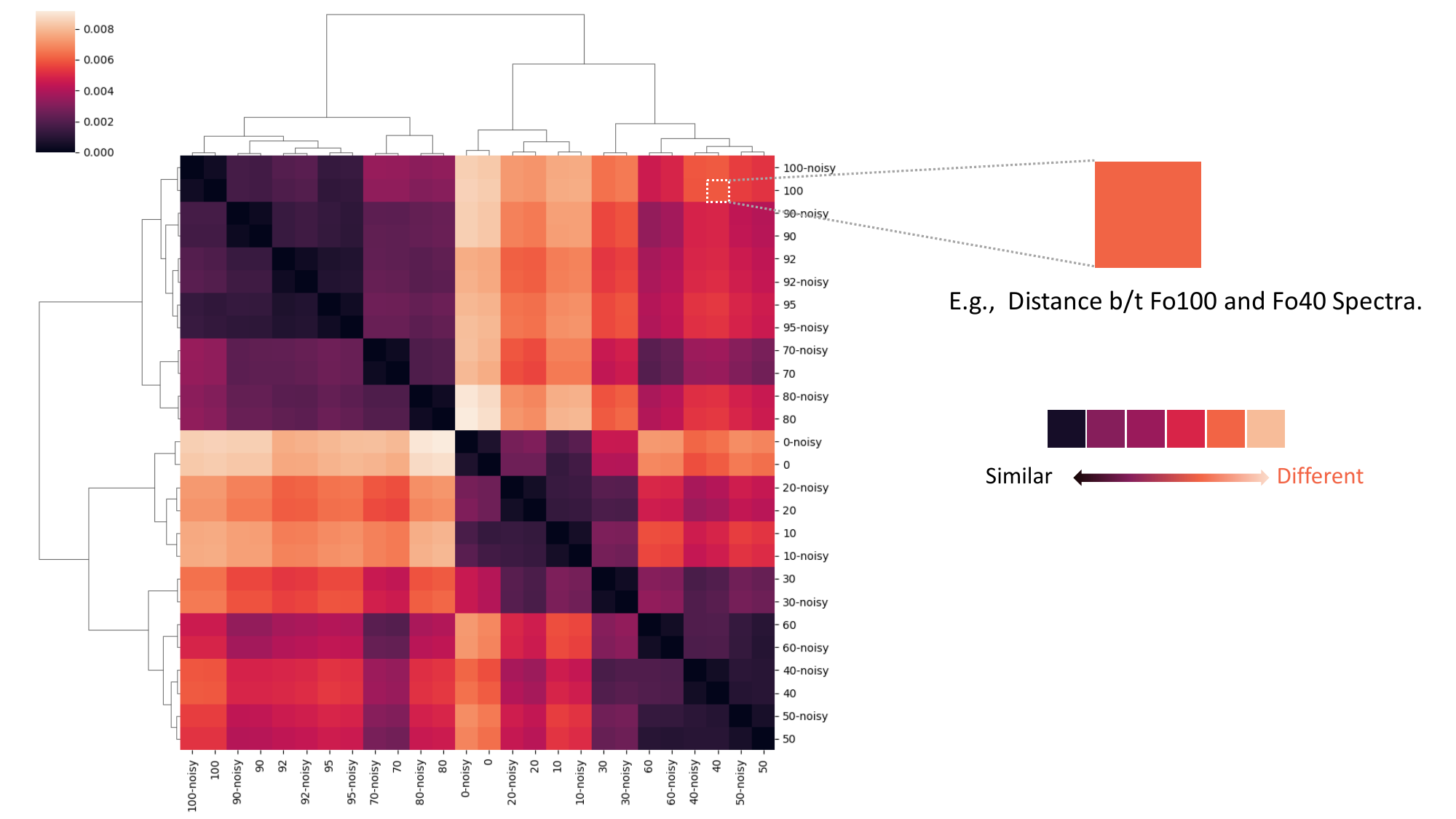}
    \caption{A Distance Matrix with EMD metric, sorted with Hierarchical Clustering. Every row and column represents a unique spectra. The right and bottom axes show the Fo number of each spectrum. The color denotes a distance score, where a darker color means the two spectra are more similar and a brighter color means they are more different from each other. The left axis dendrogram represents the clustering results. }
    \label{fig:distmatrix}
\end{figure*}

\texttt{Sequencer} outputs a distance matrix that contains distances between every two spectra, and hence is N by N in dimension as shown in Figure~\ref{fig:distmatrix}. In Figure~\ref{fig:distmatrix}, on the horizontal axis, every tick represents a library forsterite spectra with their Fo number. The matrix is symmetric, and therefore, the vertical axis to the right shows the same information. For each grid, a distance is calculated for the corresponding 2 spectra. In Figure~\ref{fig:distmatrix}, we show an example of how to read this plot. The square with a dotted white borderline is enlarged to the right. The color in the square grid maps to a distance, where a darker color represents more similarities and lighter color represents more differences. For instance, the diagonal of the matrix is black because every spectrum is being compared to itself, producing a distance of 0. 

We use a hierarchical clustering tool to cluster the distance matrix to show respective groupings. The rows in this matrix are ordered by ``Ward'' distance, where the variance between two distances is minimized as the following 

\begin{equation}
    d_{{i,j},{m,n}} = ||d_{i,j} - d_{m, n}||^2,
\end{equation}
where $d_{i,j}$ is the distance between $i^{th}$ ($j^{th}$) spectrum, and $d_{m,n}$ follows the same rule. At the initial step, all clusters are singletons, which are clusters containing a single point. Then the Ward method runs recursively to minimize the variance within each cluster. According to the given distances matrix, the row rearrangement clusters the spectra into subgroups, as visualized by the dendrogram on the left/top vertical of the distance matrix in Figure~\ref{fig:distmatrix}. The dendrograms show hierarchical, tree-like structures, in which every branch is called a ``clade'' and the terminal end of each ``clade'' is called a leaf. 

The clusters in the forsterite library from the bottom up reveal a trend of increasing Fo number that corresponds to a systematic shift in emission features towards shorter wavelengths which agrees with parametric analyses based on laboratory measurements \citep{kuebler2006, Fabian+01}, demonstrating that this non-parametric clustering methodology can be useful to discover new trends in our IRS data. In the dendrogram to the top of Figure~\ref{fig:distmatrix}, the bottom-most clades group every pair of ``clean'' and ``noisy'' with the same Fo number together, meaning that the hierarchical clustering is robust against $5\%$ of spectral noise. When we move up one level from the bottom-most clades, we see spectra with their respective nearest Fo numbers are grouped together. For example, Fa (Fo0) spectra and Fo10 spectra are grouped together, and the same applies to the clade of Fo30 and F40, the clade of Fo50 and Fo60, etc. As the hierarchical grouping with Ward Distance fully recovers the trends in Fo number, this experiment establishes that we should keep exploring the IRS dataset with this methodology to understand the composition trends in real data. 

We use the silhouette score as a metric to calculate the goodness of a clustering technique and determine the number of optimal clusters. Silhouette score has a range of [-1, 1] and is commonly used to study the separation distance between the resulting clusters \citep{shahapure2020cluster}. In essence, the silhouette plot displays a measure of how close each point in one cluster is to points in the neighboring clusters and thus provides a way to assess the number of clusters visually. 

\subsection{Visualize Sequencing Results with MDS}\label{sec:MDS_vis}

While we demonstrate that the combination of \sequencer distance matrix and hierarchical clustering is an effective approach to classify debris disk spectra according to their compositions, we also need an effective visualization tool to map the N by N, high-dimensional distance matrix onto a lower dimensional space. A low-dimensional visualization can allow us to test physically motivated correlations between the clustered spectra and their external stellar parameters (e.g., stellar luminosity, binarity, age, etc). 

We can visualize the pairwise distances in a distance matrix by using Multi-Dimensional Scaling (MDS). MDS is a commonly used statistical method to map information about the pairwise distances amongst a set of $N$ objects onto an abstract Cartesian space \citep{Mead1992}. There are 3 different types of MDS: classical, metric, and non-metric. In our case, metric MDS (mMDS) is best suited to our purpose for 2 reasons: (1). mMDS is suitable for a weighted distance matrix over iterative methods, while the classical method lacks. (2). classical MDS is strictly linear while mMDS is non-linear. Principal component analysis (PCA) is also a popular dimensional reduction algorithm that is linear. However, we chose MDS for its non-linearity over PCA linearity. We calculate the coordinates by optimizing the stress function iteratively: 
\begin{equation}
    \text{Stress}(X_1, X_2, ..., X_N)
    = \sqrt{\sum_{i=1}^N\sum_{j = 1, j\neq i}^N (d_{i,j}-||X_i - X_j||)^2},
\end{equation}
where  $X_1, X_2, ..., X_N$ are the new coordinates for each spectrum in the lower dimensional space. $d_{i,j}$ is corresponding entry in the input distance matrix for the $i^{th}$ and $j^{th}$ spectra. We minimize the stress with 300 iterations. We use \texttt{sklearn.manifold.MDS} and use the default tolerance value where eps=$1e^{-5}$.

We conduct experiments to test whether low dimensional (e.g.,  2D or 3D) MDS coordinates can produce a satisfactory representation of the distance matrix. MDS provides a set of coordinates that we can characterize with the Euclidean distance. To accurately represent N spectra, as many as (N-1) dimensions are needed. Our objective is to find as few dimensions as possible that still reproduce the original matrix reasonably. 

\begin{figure}[ht!]
    \epsscale{1.3}
    \plotone{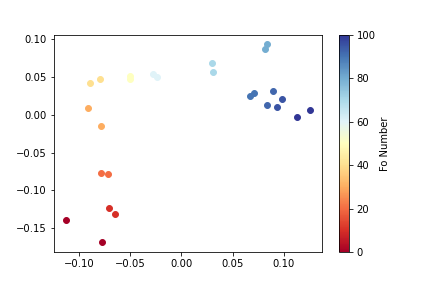}
    \caption{A 2D MDS representation of the Distance Matrix. Every circle represents a unique forsterite spectra. The colors shows their Fo number. The x and y axes are 2 arbitrary principle components that MDS generates.}
    \label{fig:MDS_folib}
\end{figure}

We find that 2D MDS generates a satisfactory low-dimensional representation of the distance matrix, because it recovers the original trends in the input data and has a small quantitative difference from the original distance matrix. In Figure~\ref{fig:MDS_folib}, we show a 2D MDS visualization where every circle represents a unique forsterite library spectra colored according to their Fo number. The x and y axes are 2 arbitrary principle components that MDS generates. We can see that the points span a clean sequence ordered by Fo number, from the top right blueish points (high Fo spectra) through the top-left yellowish (medium Fo spectra) points to the bottom-left reddish (low Fo spectra) spectra. Additionally, the relative distance between points on the plot shows the difference between Fo numbers. For example, the Fo100 spectra have a larger distance to any Fo0 spectra than to any Fo50 spectra in this cartesian coordinate.  

We quantify the absolute difference between 2D MDS presentations and the \sequencer distance matrix with statistical methods. For all the points in Figure~\ref{fig:MDS_folib}, we calculate the pairwise distance for all the points and generate a cartesian-coordinate-based distance matrix based on MDS representation. If the MDS distance matrix contains all the information from the original inputted Sequencer distance matrix, then the difference between the 2 matrices at every $(i, j)^{th}$ entry should be 0. We show such a comparison for the 2D MDS case. 
\begin{figure}[ht!]
    \epsscale{1.3}
    \plotone{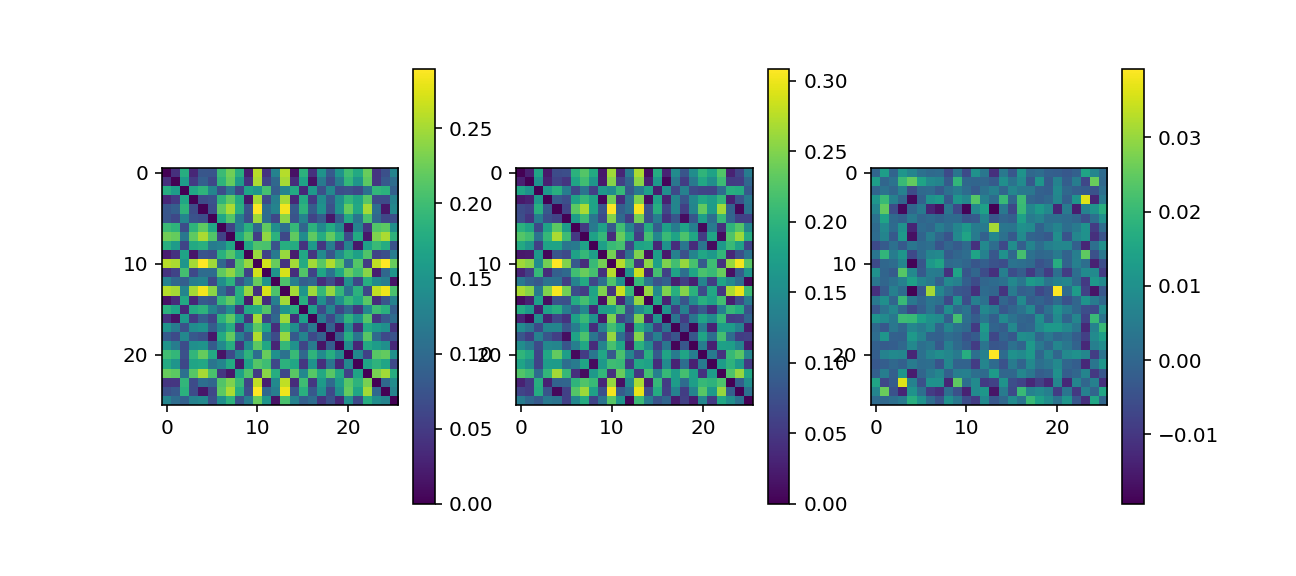}
    \caption{The comparison between Sequencer distance matrix and the MDS distance matrix. We quantify the difference between the two methods. Left: the original output of Sequencer distance matrix. Middle: The 2D MDS Euclidean Distance Matrix. Right: The difference between the two matrices.}
    \label{fig:MDS_stats}
\end{figure}
\begin{figure}[ht!]
    \epsscale{1.2}
    \plotone{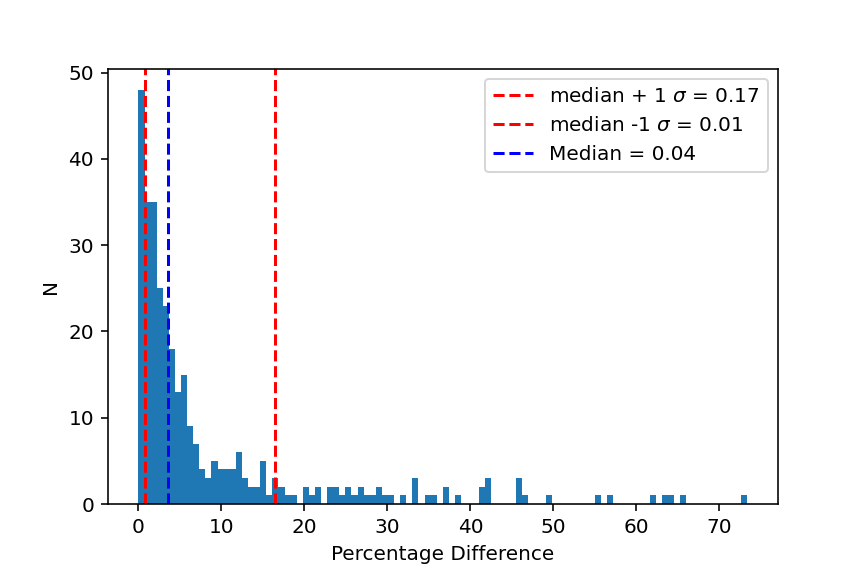}
    \caption{A histogram of the percentage difference between the MDS-based distance matrix and the original sequencer distance matrix. The percentage difference for each entry is calculated using equation \ref{eqn-diff}. The y-axis shows the cumulative number for each bin and the x-axis shows the percentage difference in units of \%. The dotted blue line shows the location of the median amongst the calculated difference values and the dotted red lines show the location of $16^{th}$ and $84^{th}$ percentile. 
    }
    \label{fig:MDS_stats_hist}
\end{figure} 
In Figure~\ref{fig:MDS_stats}, the left panel shows the original (unclustered) \sequencer distance matrix, the middle panel shows the 2D MDS based distance matrix, and the right panel shows the difference between the left and middle panels. We can see that the 2D MDS preserves the structure of the \sequencer distance matrix. We quantify the difference using 
\begin{equation}
    \text{Difference}_{i,j} = \frac{|d^{Sequencer}_{i,j} - d^{MDS}_{i,j}|}{d^{Sequencer}_{i,j}}
    \label{eqn-diff}
\end{equation}
The percentage difference can be calculated for each entry in the matrix and we show the distribution of the percentage difference in a histogram in Figure~\ref{fig:MDS_stats_hist}. We can see that the median difference is $\sim4\%$ and $16^{th}$ and $84^{th}$ is ($1$, $17$)\%. As such, the MDS 2D representations reduces the original $24$ x $24$ parameter space into a 2D representation and only incurs a median of $\sim 4\%$ error to each grid.  
% In the following sections, we employ MDS to explore the correlation between stellar parameters and the grouping of spectra.  

\subsection{Purposes of CLUES}
To give a high-level summary of the tool, CLUES introduces two main ways to understand relationships within a dataset: identifying clusters and uncovering intrinsic trends. The former combines an agglomerative hierarchical clustering algorithm and multi-dimensional scaling to effectively identify clusters within a high-dimensional dataset and quickly narrow down the vast parameter spaces for subsequent modeling. The latter is directly inherited from \sequencer, which aims to find a maximum spanning linear trend within the data, known as MST. Unlike clustering which maximizes density within clusters, MST maximizes sparseness. One could not use MST to directly identify clusters based on the proximity of nodes because MST is non-unique, and the concept of ''cluster" is not well-defined in trees in general \citep{yu2015, habib2022}. We further discuss the usages of these two approaches with three examples detailed in the next section (Sect. \ref{section:results}). Amongst the three examples, example 2 -- IRS spectrum of HD 113766 -- well-exemplifies the differences between the two approaches and the need for agglomerative clustering rather than MST only.

\section{Results} \label{section:results}
In this section, we run three experiments using data relevant in astrophysical (planet formation) and geological context to illustrate the utility of the \texttt{CLUES} workflow. The three examples are as below: \\
(1). \textbf{A library of 34 mineral emissivity spectra}: The ingredients are relatively pure minerals drawn from Earth's (and other terrestrial planets') crustal and mantle compositions. For example, quartz and calcite are selected as exemplars of terrestrial planet crustal minerals, forsterite and enstatite are selected as exemplars of Earth mantle minerals. These minerals in the library include forsterite, enstatite, aragonite, orthoclase, quartz, microcline, dolomite, calcite and anorthite and are collected from the JPL ECOSTRESS spectral library and the JENA database \citep[][]{ecostress, Jager+03, Henning+99}. The exact species used are listed in Table \ref{tbl-species}. Our goal is to examine whether the \texttt{CLUES} workflow can robustly classify different classes of minerals non-parametrically. \\ %For the minerals, write down the chemical formula here also for easier understanding .
(2). \textbf{One debris disk spectrum + 34 Mono-mineralic Spectra in the Emissivity library}: We take one step further away from solar system minor bodies to using remotely-sensed mid-infrared spectra of an exoplanetary system HD 113766 taken with \textit{Spitzer/IRS}. We shuffle HD 113766's spectrum amongst the library emissivity spectra as described in (1). Our goal is to test whether the \texttt{CLUES} workflow can determine the dominant mineralogy for the debris disk and compare this with the detailed work done on this system \citep{Lisse+08, Olofsson+12}. This example is a step towards a full debris disk spectral analysis in Paper II.

(3). \textbf{An ensemble of 59 meteorite spectra}: These chondritic and achondritic meteorite samples are mixtures of various minerals and therefore more complex in their compositions from laboratory-forged or mineral collected from Earth's surface. The meteorites are tangible samples of minor bodies originated from the debris disk of our solar system \citep[as seen in our solar system that minor bodies are from main belt, e.g., ][]{Binzel2015, moskovitz2010, Menichella+96} following planetary surface evolution since the epoch of solar system formation to present. For the spectral measurements, the meteorites are ground up into fine particles before reflectance measurements are taken in labs. We use them to test whether mixed compositions in minor bodies can be classified by our \texttt{CLUES} workflow.
\begin{deluxetable}{lcc}
    \tablecaption{Mineral Species Optical Constant References}\label{tbl-species}
    \tablehead{
    \colhead{Species}& \colhead{Grain Properties
    } & \colhead{Ref \& Notes} }
    \startdata
    Forsterite & $2\mu$m & Jena, (1) \\
    Enstatite & $2\mu$m &  Jena, (1) \\
    Quartz (Si$\mathrm{O_{2}}$) & Fine, $<2\mu$m  & \href{https://speclib.jpl.nasa.gov/ecospeclibdata/mineral.silicate.tectosilicate.fine.tir.quartz_1.jhu.nicolet.spectrum.txt}{Data Access}, (2)\\
    Anorthite ($\mathrm{CaAl_{2}Si_{2}O_{8}}$)   &  Fine, $<2\mu$m  & \href{https://speclib.jpl.nasa.gov/ecospeclibdata/mineral.silicate.tectosilicate.fine.tir.anorthite_1.jhu.nicolet.spectrum.txt}{Data Access}\\
    Orthoclase ($\mathrm{KAlSi_{3}O_{8}}$)  &  Fine, $<2\mu$m  & \href{https://speclib.jpl.nasa.gov/ecospeclibdata/mineral.silicate.tectosilicate.fine.tir.orthocl_3.jhu.nicolet.spectrum.txt}{Data Access}\\
    Microcline ($\mathrm{KAlSi_3O_8}$)  &  Fine, $<2\mu$m  & \href{https://speclib.jpl.nasa.gov/ecospeclibdata/mineral.silicate.tectosilicate.fine.tir.micro_1.jhu.nicolet.spectrum.txt}{Data Access}\\
    Albite ($\mathrm{NaAlSi_3O_8}$)& Fine, $<2\mu$m    & \href{https://speclib.jpl.nasa.gov/ecospeclibdata/mineral.silicate.tectosilicate.fine.tir.albite_1.jhu.nicolet.spectrum.txt}{Data Access}\\
    Dolomite ($\mathrm{CaMg(CO_3)_2}$)  & Fine, $<2\mu$m    & \href{https://speclib.jpl.nasa.gov/ecospeclibdata/mineral.carbonate.none.fine.tir.dolomite_1.jhu.nicolet.spectrum.txt}{Data Access}\\
    Calcite ($\mathrm{CaCO_3}$) & Fine, $<2\mu$m    & \href{https://speclib.jpl.nasa.gov/ecospeclibdata/mineral.carbonate.none.fine.tir.calcite_2.jhu.nicolet.spectrum.txt}{Data Access}\\
    Aragonite ($\mathrm{CaCO_3}$)&  Fine, $<2\mu$m  & \href{https://speclib.jpl.nasa.gov/ecospeclibdata/mineral.carbonate.none.fine.tir.aragonite_1.jhu.nicolet.spectrum.txt}{Data Access}\\
    \enddata
    \tablecomments{(1). See \href{https://www.astro.uni-jena.de/Laboratory/OCDB/crsilicates.html}{Jena Database for crystalline silicates optical constants}. (2). Data from NASA/JPL ECOSTRESS Spectral library. } 
    \end{deluxetable}
\begin{figure*}[ht!]
    \epsscale{1.1}
    \plotone{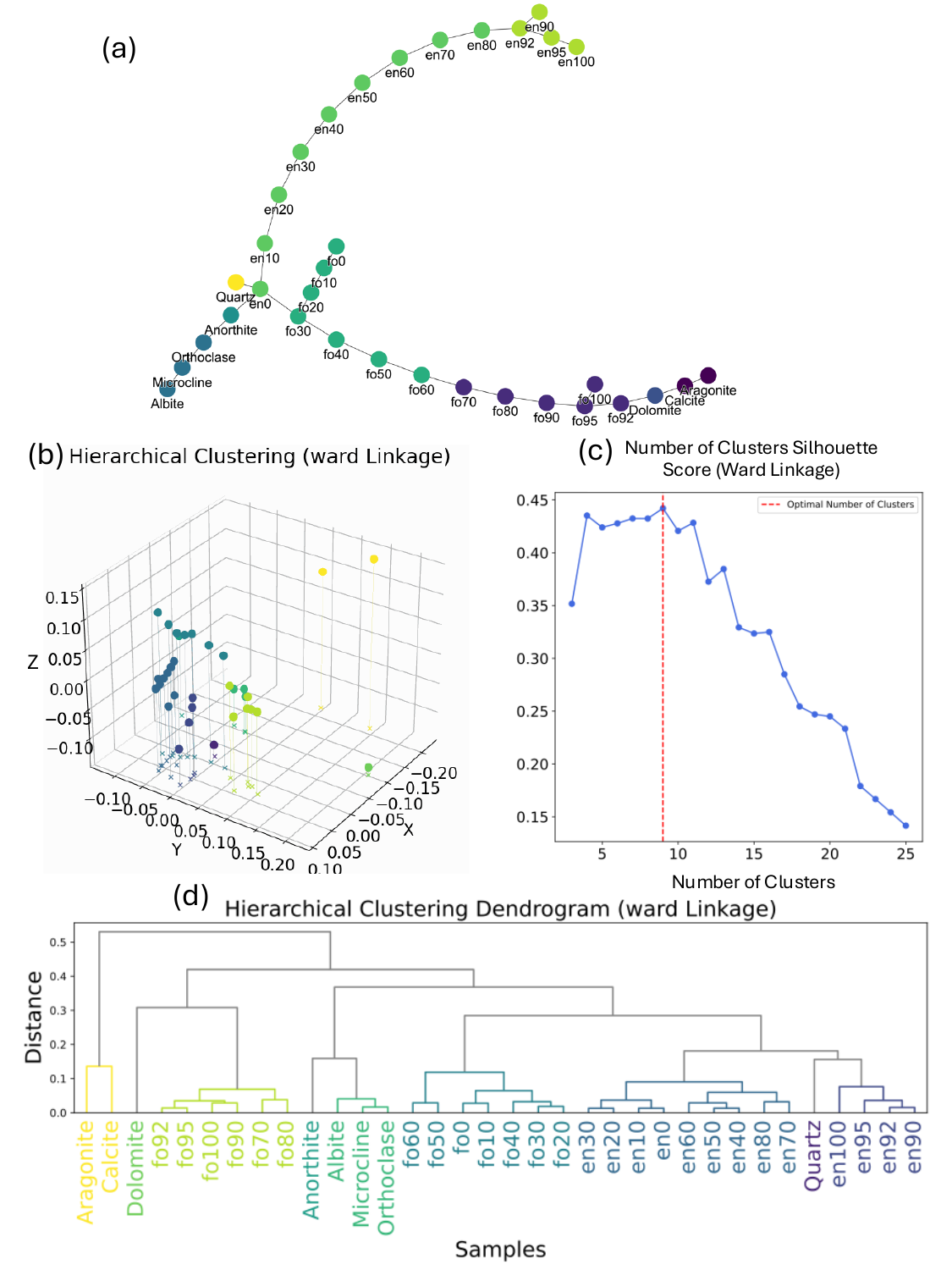}
    \caption{CLUES Output for a subset of ECOSTRESS Spectral library spectra. \textbf{(a)}.Minimum Spanning Trees (MST) computed using \texttt{Sequencer}. \textbf{(b).} Using 3D-MDS to visualize distance matrix. \textbf{(c).} Silhouette Score for Determining the optimal number of clusters. \textbf{(d).} Dendrogram visualization for clustering a library of mineral spectra using Ward linkage.}
    \label{fig:clues-d1}
\end{figure*}
\begin{figure*}[ht!]
    \epsscale{1.05}    
    \plotone{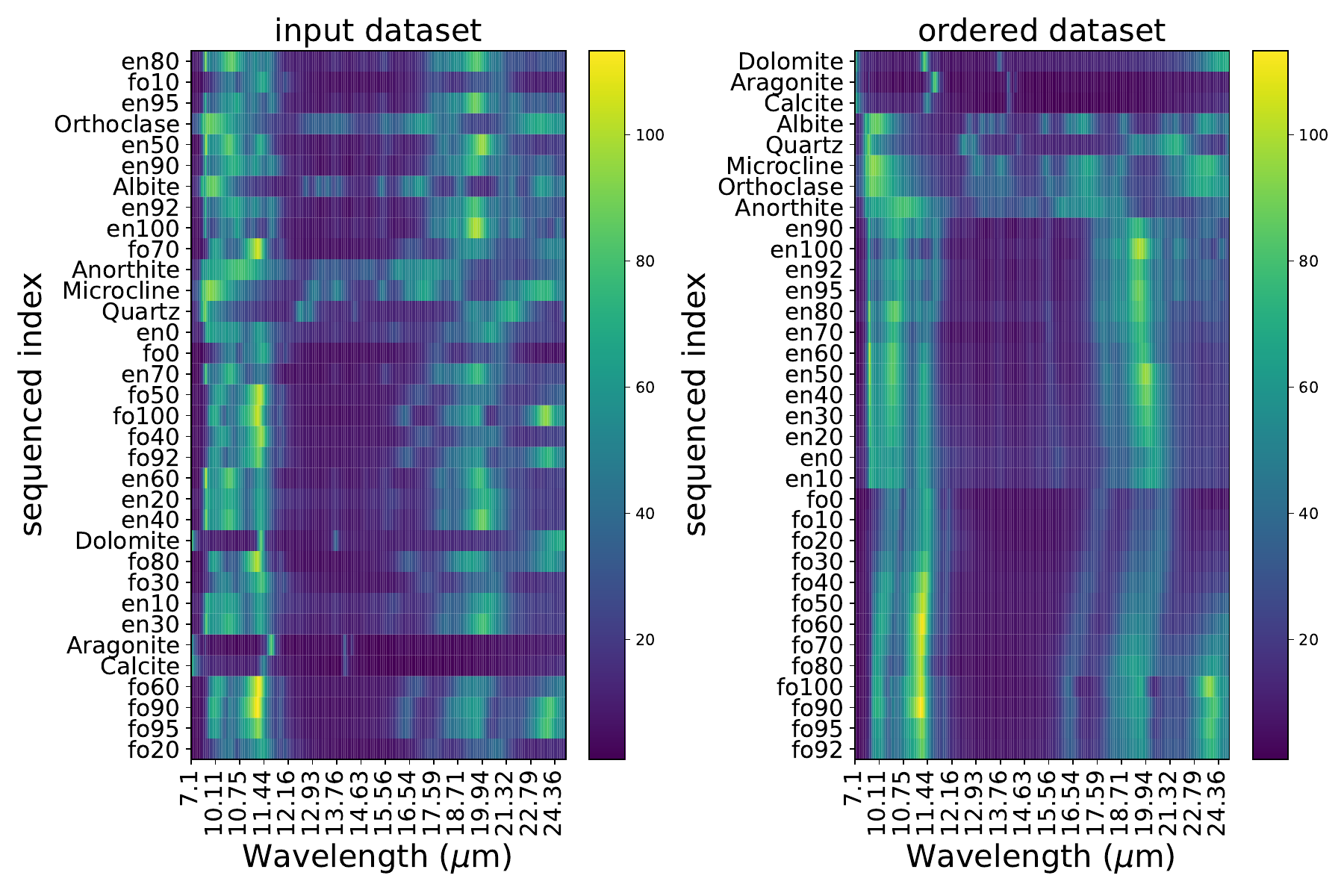}
    \caption{1D Sequence from the MST of the Emissivity Library using a distance scale of 10. The left-hand side shows the input spectra while the right-hand panel shows the ordered spectra.}
    \label{fig:mst1dseq}
\end{figure*}
\begin{figure*}[ht!]
    \epsscale{1.05}
    \plotone{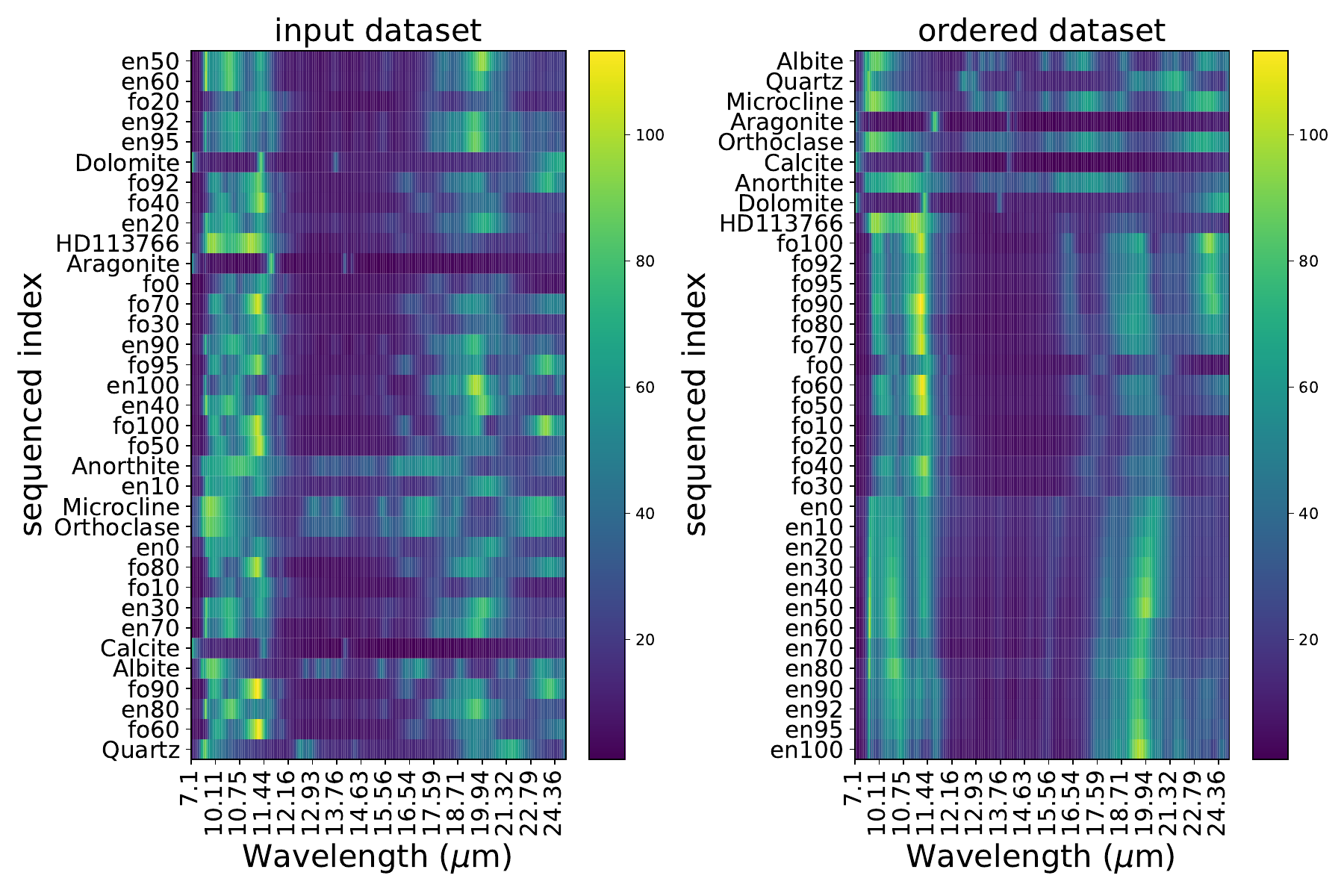}
    \caption{1D Sequence of MST including a debris disk spectra with a distance scale of 15. }
    \label{fig:mst1dseq-irs}
\end{figure*}
\begin{figure*}[ht!]
    \epsscale{1.1}
    \plotone{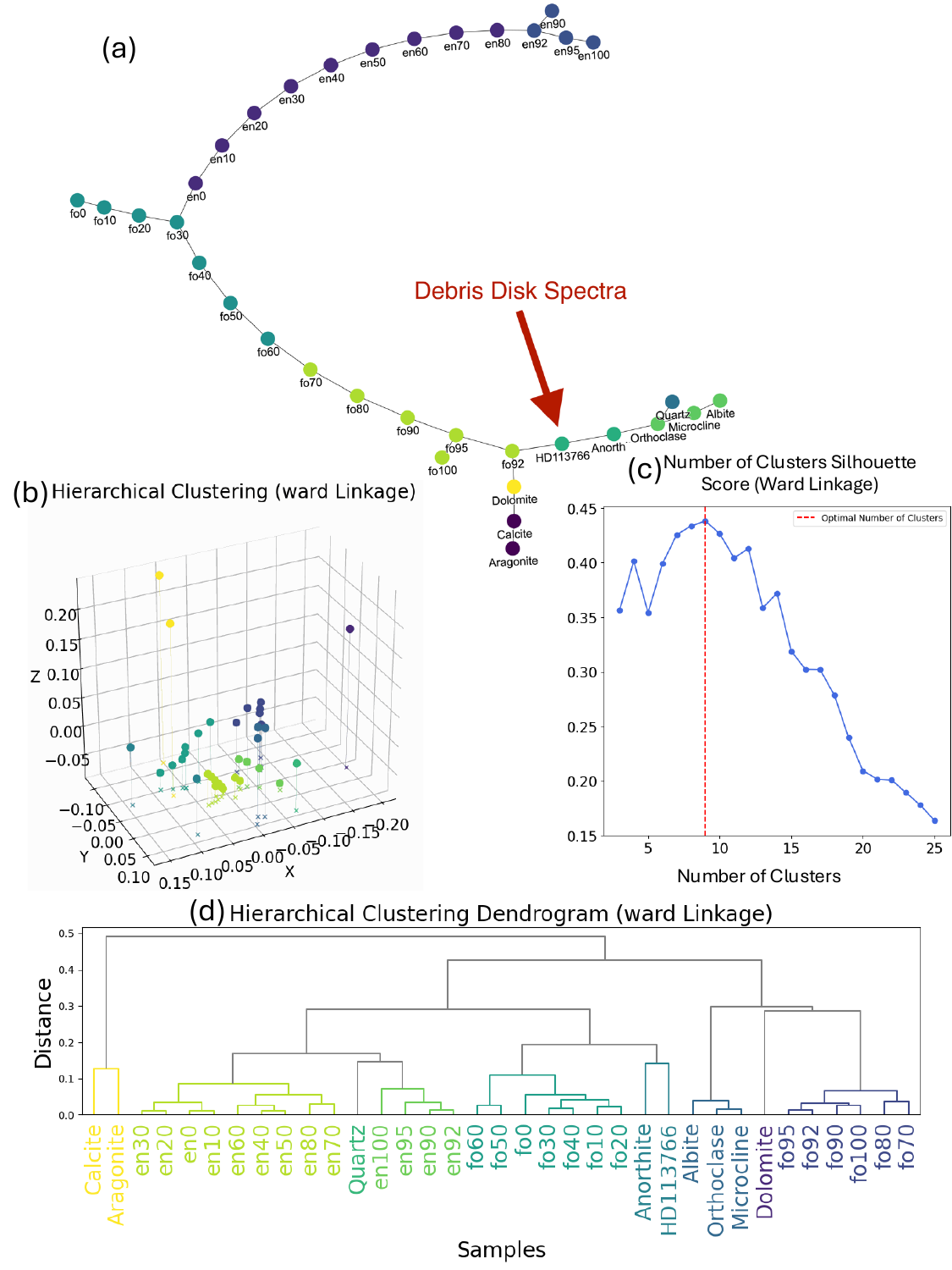}
    \caption{\texttt{CLUES} Output for a debris disk spectrum mixed amongst a subset of ECOSTRESS Spectral library spectra. \textbf{(a)}. Minimum Spanning Trees (MST) computed using \texttt{Sequencer}. \textbf{(b).} Using 3D-MDS to visualize distance matrix. \textbf{(c).} Silhouette Score for Determining the optimal number of clusters. \textbf{(d).} Dendrogram visualization for clustering a library of mineral spectra using Ward linkage.}
    \label{fig:clues-d2}
\end{figure*}
\begin{figure*}[ht!]
    \epsscale{1.2}
    \plotone{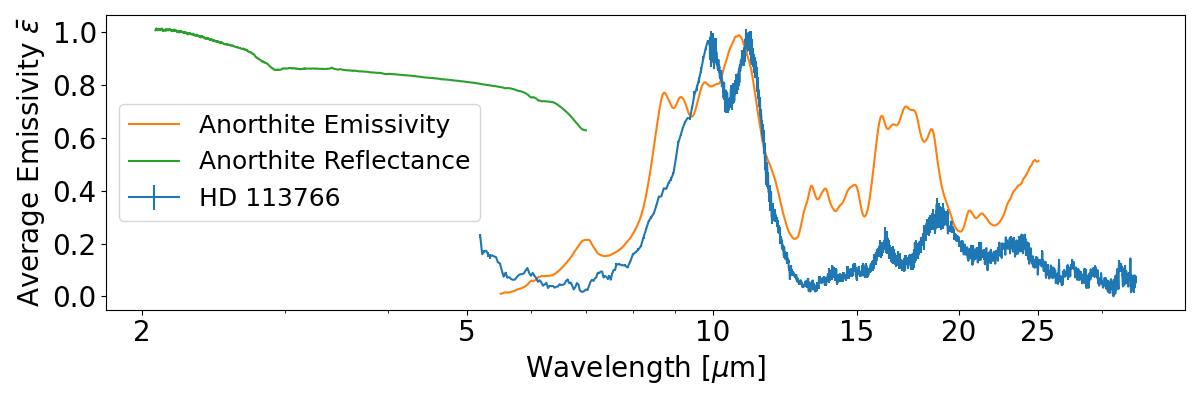}
    \caption{Comparison between HD 113766 Average Emissivity spectra and Anorthite Fine Grain Emissivity Spectrum}
    \label{fig:anorthite}
\end{figure*}
\begin{figure*}[ht!]
    \epsscale{1.1}
    \plotone{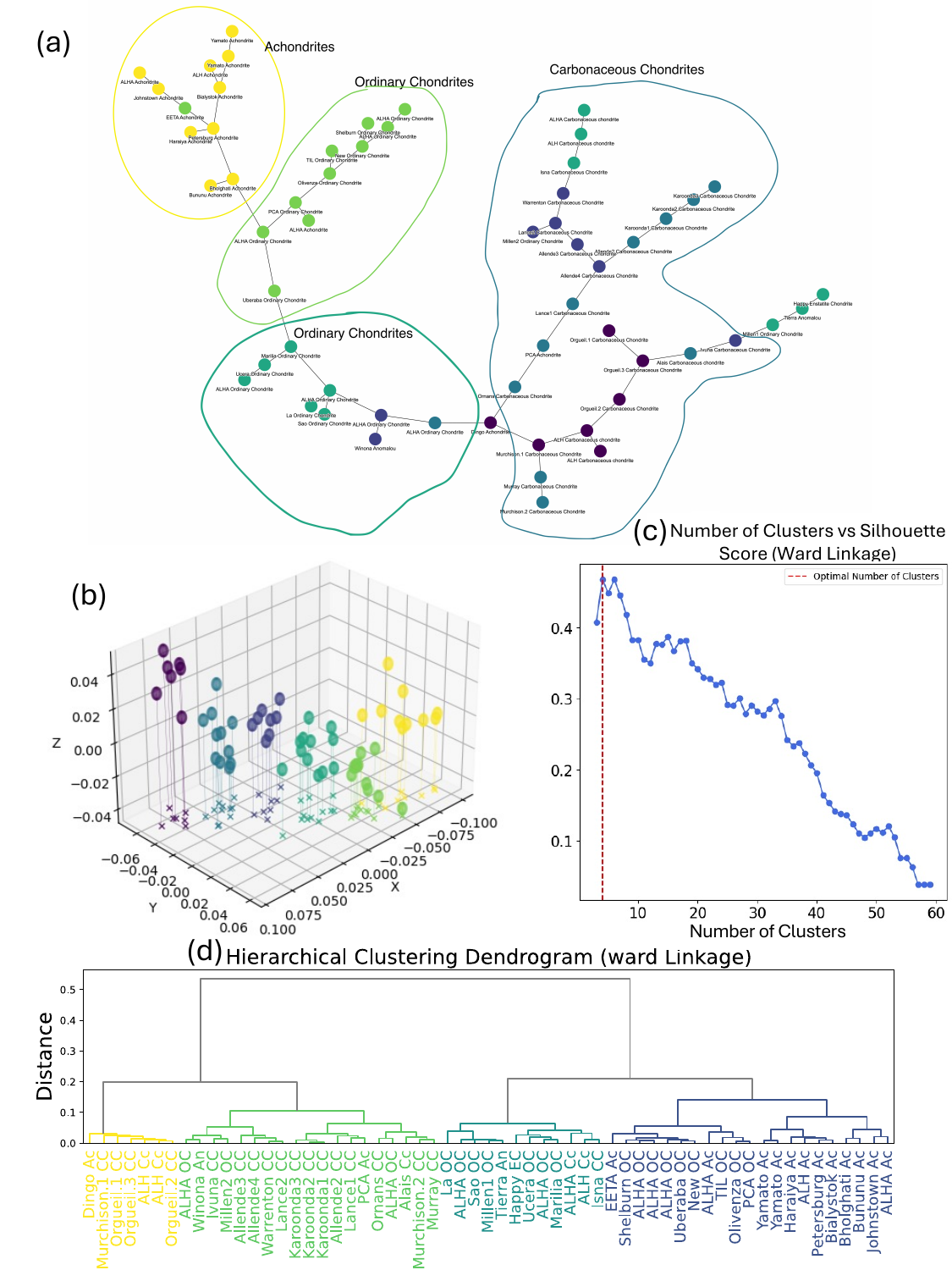}
    \caption{\texttt{CLUES} Output for 59 meteorite spectra from ECOSTRESS Spectral library. \textbf{(a)}.Minimum Spanning Trees (MST) computed using \texttt{Sequencer}.  The annotations serve as visual guides to various groups of meteorite spectra according to the hierarchical clustering results. \textbf{(b).} Using 3D-MDS to visualize distance matrix. \textbf{(c).} Silhouette Score for Determining the optimal number of clusters. \textbf{(d).} Dendrogram visualization for clustering a library of mineral spectra using Ward linkage.}
    \label{fig:clues-d3}
\end{figure*}
\subsection{Non-parametric Clustering Results}
% \begin{figure}[ht!]
%     \epsscale{1.0}
%     \plotone{figures/SparseDM_EMD.pdf}
%     \caption{Sparse Distance Matrix with Hierarchical Clustering. The Sparse DM is calculated using EMD.}
%     \label{fig:mst1dseq}
% \end{figure}
In this section, we present results from non-parametric clustering using \texttt{CLUES} with three ensembles of spectra, (1) the monomineralic spectral library, (2) one (polymineralic) debris disk spectrum mixed in the monomineralic spectral library, and (3) the polymineralic meteorite spectral library. These three datasets are increasingly complex in their composition, and therefore are good examples to test the effectiveness and limitations of our algorithms.

\subsubsection{Mineral Spectral Library}\label{results:lib1}
We construct a common wavelength axis for the ECOSTRESS library spectra, Jena library spectra, and the IRS spectra. The majority of the mineral spectra in the ECOSTRESS spectral library cover a wavelength range from $2.1$ to $25\,\mu$m. Although the IRS wavelength spans from $5.1$ to $37\,\mu$m, we truncate the $5.1$--$7.1\,\mu$m and $33$--$37\,\mu$m to minimize the effects of spectral point-to-point calibration issue and fringing, as discussed in sections \ref{sub:norm} and \ref{section:analysis}. After truncation, we preserve the original IRS wavelength grid, and down-sampled the higher-resolution (R $\sim$ 1000-3000) library spectra to the corresponding IRS wavelengths. The resulting library spans from $7.1$\,--\,$24.36\,\mu$m and has a resolution of R$\sim$300. For a larger IRS debris disk dataset mainly consists of low-resolution data in Paper II, the common wavelength axis would require a lower resolution of R$\sim\,60$\,--\,$100$.

For the library of Earth-based mineral emissivity spectra (plotted in Figure~\ref{fig:clues-d1} - Panel a), we start with the \sequencer algorithm that helps us find the optimal elongation of the MST after comparing a number of different scales and distance metrics. We find that a scale of 10 with EMD distance metric gives a maximum elongation of MST and thus this contributes primarily to the distance matrix. One of the most immediate low-dimension visualizations for MST is a collapsed 1D sequence. In Figure~\ref{fig:mst1dseq}, we show the example of the 1D sequenced spectra, where each spectrum is sequenced with respect to their spectral features. Although 1D sequence omits information such as elongation and cluster information that are originally contained in MST, the advantage of 1D sequence is that it gives a very direct overview of how spectral features shift in the wavelength space as a result of mineral intrinsic properties such as Fe/Mg ratio (stoichiometry) and different mineral species. However, as we can see in Figure~\ref{fig:mst1dseq}, the 1D sequence doesn't nicely order all the spectra in our datasets. Thus, we need the full \texttt{CLUES} analysis framework for clustering the spectra and finding exemplars. 

We start by visualizing the MST (Figure~\ref{fig:clues-d1} - panel a) where each node represents a unique spectrum in our library. MST visualization emphasizes the difference between enstatite and forsterite by separating them into 2 different branches and states that spectra of non-silicate minerals such as calcite and aragonite are more similar to olivine end-members than that of pyroxene end-members, enstatite. The feldspar minerals (aluminium tectosilicate minerals; e.g., Anorthite, Orthoclase, Albite) that make up a large fraction of the continental crust, form a distinct branch in the MST. 

An alternative approach for clustering is using the \sequencer algorithm results to obtain the optimally weighted distance matrix. We show dendrograms resulting from hierarchical clustering with ward linkage in Figure~\ref{fig:clues-d1} panel (d). These results show that at the highest level (top of the graph) of the dendrogram, the spectra are classified into 2 main branches, a ``non-silicate'' mineral class that contains calcite, and aragonite and a ``silicate'' class that contains forsterite, enstatite, feldspars, and quartz as well as dolomite. While dolomite is not a silicate, it is grouped as an outlier with the silicate group (Figure~\ref{fig:clues-d1} panel d) and is close to the carbonates overall as seen clearly in the MST. In the silicate class, spectra are further divided into the forsterite group, enstatite group, and feldspar group. As both the quartz and enstatite group have spectral features at $12$ and $15\,\mu$m, they are grouped closer to each other compared to forsterite, which doesn't have any feature between $12$ to $16\,\mu$m. When we dive deeper down, we find that even within a group wherein the composition is broadly similar, the hierarchical clustering can further distinguish the difference between high-Fe ratio and low-Fe ratio stoichiometry (known to geologists as end-members) in forsterite and enstatite. While 
low Mg/Fe spectra such as Fo0, Fo10, Fo20, Fo30, Fo40 are grouped together (see the bottom level clades in Figure~\ref{fig:clues-d1} panel d), similarly high Mg/Fe spectra Fo100, Fo92, and Fo90 are grouped into a single cluster. The silhouette score analysis for hierarchical clustering suggests that the optimal number of groupings is 9 groups (with each group having a distinct color) - this clearly separates the high vs low Mg/Fe forsterite and 
enstatite, feldspars, and carbonate groups.

We can compare the results of the two approaches by coloring the nodes in the MST by their respective groupings from hierarchical clustering based on the distance matrix. In essence, Figure~\ref{fig:clues-d1} showcases the possibility of simultaneously comparing 2 different ways of grouping based on the distance matrix. Broadly, we find the results are consistent with the two approaches, thus providing further confidence in the overall results as well as illustrating the utility of the \texttt{CLUES} methodology.

\begin{deluxetable*}{lccccc}
    \tablecaption{Comparison of Results amongst 3 Experiments/Datasets}\label{tbl-MST-results}
    \tablehead{
    \colhead{Dataset}& \colhead{Metric} & \colhead{Scale (l)
    } & \colhead{Elongation} & \colhead{Hierarical Clustering Linkage}& \colhead{Optimal Cluster No.}}
    \startdata
    Lab & EMD & $10$ & $23.61$ & Ward &$9$ \\
    IRS+Lab & EMD & $15$  & $21.2$& Ward &$9$\\
    Meteorite & EMD & $1$ & $22.1$ & Ward & $4$ or $6$\\
    \enddata
    % \tablecomments{(a). } 
    \end{deluxetable*}

\subsubsection{Classifying a Debris Disk spectrum using a Mineral Spectral Library}\label{results:lib2}
Next, we add a remotely-sensed debris disk spectrum to the library from the previous section and explore and understand its composition in relationship with the pure mineral spectra. The debris disk of choice is HD 113766 because its composition is well-studied with MIR spectra \citep{Lisse+08, Olofsson+12}. IRS spectra of HD 113766 include SL ($5.2$\,--\,$14.5\,\mu$m, R$\,\sim\,$100), SH ($9.9$\,--\,$19.6\,\mu$m, R$\,\sim\,$600) and LH ($18.7$\,--\,$37.2\,\mu$m, R$\,\sim\,$600) modes. We use the SL for $5.2-10\,\mu$m region and high-resolution mode from $10$\,--\,$37.2\mu$m region, resulting in a spectral resolution of R\,$\sim\,300$. Although the resulting spectrum spans from $5.2$\,--\,$37.2\mu$m, we use $7.1$\,--\,$24.36\,\mu$m region to minimize the effects of spectral point-to-point calibration issue, fringing and for constructing common wavelength grid, as discussed in sections \ref{sub:norm}, \ref{section:analysis} and \ref{results:lib1}.

We re-run our \texttt{CLUES} analyses with the addition of debris disk spectra and show our results in Figure~\ref{fig:clues-d2}. Figure~\ref{fig:clues-d2} panel (a) shows a similar MST as Figure~\ref{fig:clues-d1} panel (a) but now with HD 113766 surrounded by spectral library dataset. Although MST cannot be used to interpret the clustering of the spectra, MST reveals new possibilities in the composition parameter space. MST shows HD 113766's composition is close to Forsterite. Previous spectral modeling work on this target shows that relatively Fe-rich forsterite (Fo80) is present \citep{Olofsson+12} as opposed to Mg-rich Forsterite (Fo100). Clustering results in panel (d) are consistent with the detailed fitting, where HD 113766 is preferentially grouped with forsterite enriched in Fe content (cyan group, Fo 60 to Fo0) as opposed to other mineral species. The CLUES algorithm also offers new insights by pointing to potential contribution of anorthite (a common igneous mineral) in the HD113766 spectra - this is a mineral species that has not been previously considered in disk analysis. In the case of HD 113766, anorthite is clearly not the only dominant mineral species since although its emission feature shows a highly similar FWHM in the $10\,\mu$m to the disk data, there are significant difference in the $20\,\mu$m features (Figure \ref{fig:anorthite}). Although in this specific case, anorthite may not be most useful composition, its similarity in the $10\,\mu$m potentially suggests new compositions to consider for fitting the warm dust features.

The HD 113766 example also shows that for polymineralic spectra, such as that of debris disks, embedded in pure monomineralic library spectra, the clustering technique outperforms the MST-based collapsed 1D sequence. If we compare, the 1D sequence of a pure mineral library shown in Figure \ref{fig:mst1dseq} right panel with the 1D sequencer of an IRS spectrum embedded in Figure \ref{fig:mst1dseq-irs} right panel, we can see that the top rows in the later do not effectively sequence compositions from albite to dolomite, while the former does. This is due to the fact that HD 113766 spectrum is a composite spectrum with various compositions and its relation with either the silicate group or the other group cannot be adequately represented on a 1D sequence. 

In addition, trends in linear spaces show that the MDS-reduced embedding representations produce a qualitative similar relationship in a simple 3D cartesian space that allows hierarchical clustering to perform well. Figure~\ref{fig:clues-d2} panel b shows the 3D MDS embedding overlaid with colors showing hierarchical clustering with ward linkage. This is particularly helpful in analyzing data that has thousands of dimensions such as a high-resolution IFU data cube that can be computationally intensive. Performing a dimensionality reduction algorithm on those data would give us an efficient way to understand the correlation amongst features in the parameter space as a first pass.   

The results for HD 113766 also highlight an important practical consideration when using \texttt{CLUES} (or other similar clustering and unsupervised learning approaches) - the role of ancillary data such as photometry and whether/how to include them in the analysis. Taking HD 113766 as an example, one could rule out anorthite as a significant dust species by including the NIR portion of the anorthite spectrum in the analyses and comparing with the available photometry or non-IRS datasets (e.g., IRTF NIR data). In Figure~\ref{fig:anorthite}, the anorthite emissivity spectrum is overplotted on HD 113766's average emissivity spectrum over two wavelength regions, the NIR (from $2$\,--\,$5\,\mu$m), and MIR (from $7$\,--\,$25\,\mu$m). With the knowledge that HD 113766 lacks any prominent emission features in the NIR, we can rule out the anorthite composition. However, from an algorithmic perspective, if these additional constraints and even upper limits on the strength of emission features are not included in inputs, CLUES will not filter the associated minerals and they have to manually checked by the user. However, we still provide a much smaller set of target minerals compared to 1000s of possible mineral species in the spectral libraries.

\subsubsection{Meteorite Spectra}\label{results:lib3}
We now turn away from the laboratory-based, monomineralic spectral library and present results on applying \texttt{CLUES} to polymineralic samples, a library of 59 meteorites reflectance spectra with a wavelength range from $2.1$ to $25.0\,\mu$m. In Figure~\ref{fig:clues-d3}, we show the MST of all 59 meteorite spectra from the ECOSTRESS database. As annotated in Figure~\ref{fig:clues-d3} panel (a), we can see that the achondrites, carbonaceous chondrites, and ordinary chondrites separate cleanly on the MST and also by hierarchical clustering (as colors separate cleanly according to the groupings). The point of this exercise is to show that for polymineralic samples, \texttt{CLUES} allows us to quickly group spectra according to their compositions and offers direct insights into composition space with minimal time costs. 

We summarize the meta statistics of MST for our 3 datasets in Table \ref{tbl-MST-results}.  We can see that MST provides useful information such as elongation, and using the different distance measures, we can get an optimally weighted distance matrix for further non-parametric classifications in \texttt{CLUES}.

\section{Discussion} \label{section:discussion}

\subsection{Sequencer vs \texttt{CLUES} and Limitations to Our Methodology}
The \texttt{CLUES} workflow is directly built upon the \sequencer algorithm, though we add a significant amount of additional unsupervised clustering components. While the \sequencer excels at finding underlying patterns non-parametrically and calculating distance between every pair of spectra, it does not perform well on clustering tasks for spectra with broad (a few micron in FWHM) and correlated emission features and fringing noises. The lack of correlation between spectral features from different minerals demands an additional clustering algorithm that specializes in taking a distance matrix as an input and clusters spectra into compositionally distinct groups. In addition, the \sequencer is designed to find a 1D sequence of the spectra as the end product - this isn't feasible or useful for all spectral data. This is why we build \texttt{CLUES}, which utilizes the best parts of both \sequencer and hierarchical clustering algorithms. In contrast, \sequencer focuses exclusively on using MST to generate a collapsed 1D sequence that maximizes the span of any 1D trends within the data. 

\subsection{Robustness of \texttt{CLUES} and its limitation}
One of the practical challenges with real data analysis is that many Spitzer IRS disk observations have low SNR and there can be substantial fringing noise. While our data pre-processing prior to the \texttt{CLUES} analysis helps to reduce these issues, noise in the spectra remains a challenge. To analyze the sensitivity of our results, we perform \texttt{CLUES} on library spectra with simulated gaussian noise (uncorrelated between each spectral wavelength). We find that when the spectral noises reach $20$\%, \texttt{CLUES} performance starts to degrade over all scales \textit{l}. The maximum elongations with all scales start to decrease to a small number which can be a factor of 10 less than the total number of spectra in the sample. Moreover, the hierarchical clustering algorithm starts to confuse enstatite spectra with forsterite spectra because simulated random spectral noise characteristics (such as fringes) starts to dominate over bona-fide spectral features in both the feature amplitude and wavelength span. Thus, we have an empirical estimate of the SNR of $>$5 that is needed for reasonable clustering results. 

\subsection{Distinguish Mixed Composition from Pure Mineral}
\texttt{CLUES} performs well with mixed materials such as meteorites and debris disk dust grains, making it a suitable first step for classifying spectra in a non-parametric, unbiased way. Traditionally, human experts examine the spectral features based on individual prior experience and can guess the suitable compositions to add to a model. However, when the spectra consists of a large amount of components, the parametric space quickly becomes too big for human eyes. \texttt{CLUES} enables an automated and bias-free first-pass examination of spectra and provides a good guess of the parameter space as a starting point for modeling. While the mineral spectral library contained 34 components, one can include a much larger dataset as a first estimate of the relevant components in a debris disk spectrum.

In addition, \texttt{CLUES} workflow is also suitable for testing the intrinsic properties of dust grains. Even for the same composition, spectra can still be significantly affected by additional factors such as dust grain sizes, dust grain shape, and temperatures that they are emitting. Even with a few compositions, these dust property quickly expands the parameter space (e.g., with 4 compositions, the dust intrinsic properties would multiply it by a factor of 3 and make it a 12-D parameter space), making it computationally intensive. \texttt{CLUES} enables us to test these parameters by embedding a spectrum of interest in a library that consists of emissivity spectra with similar compositions but slightly varies in dust grain size distribution, shape, and temperature. While not shown in the figure, we calculate forsterite and enstatite emissivity from Jena optical constants by varying grain size diameters from $0.1$--$10~\mu$m in radius for a single-sized population. In our dataset, \texttt{CLUES} is able to distinguish the grain sizes for HD 113766 and cluster HD 113766's spectrum amongst the library spectra with the grain size distribution between $2$--$5~\mu$m. 

\subsection{Application to MIR IFU Data: the Case of \textit{JWST} MIRI/MRS IFU Data}
In addition to archival spectra from Spitzer IRS, \texttt{CLUES} will be useful for space-based MIR IFU data such as \textit{JWST}/MIRI. \textit{JWST}'s MIRI IFU spectra covers $5$ to $28~\mu$m and is sensitive to dust and gas spectral features due to thermal emission. Given that \textit{JWST}/MIRI wavelength coverage and spectral resolution (R$\sim$ $100$\,--\,$3000$) are similar to that of IRS spectra and library spectra used in our example, \texttt{CLUES} will be effective for the analyses of \textit{JWST}/MIRI spectra. 

More importantly, the complexity of MIRI IFU data increases the difficulty in analyzing the spectra using the traditional, parametric spectral modeling techniques from long-slit spectroscopy. For a target previously observed with Spitzer/IRS, the MIRI/MRS IFU observation of the same target will likely possess more complicated 2D PSFs, because of the MRS's improved spatial resolution. With appropriate preprocessing steps (such as PSF subtraction, flux normalization) to MRS IFU data, \texttt{CLUES} can also be applicable for analyzing spectral features from spatially resolved IFU data cubes. 

% Basically highlighting the utility/possibility and maybe things that would need to be changed ? I would suggest at least specifically focus on IFU spectral cases where finding representive end-members in data cubes could be useful. 

\subsection{Application to NIR IFU Data}
Similar to the application on MIR IFU but over a slight different wavelength range, \texttt{CLUES} is also useful for NIR IFU data such as space based \textit{JWST}/NIRSpec IFU and ground-based Gemini/GNIRS IFU data. Unlike in the MIR where thermal emission of dust grain dominates the flux, in the NIR wavelength, the combination of reflected light and thermal emission contribute to the observed spectra, where reflected light and thermal emission have different underlying physical processes. While not presented in this paper, the JPL laboratory measurements also include extensive reflectance spectra which could be useful for applying \texttt{CLUES} to NIR IFU data for compositional analyses.   

In addition, various distance measures can be adapted for these NIR IFU data with \texttt{CLUES}. For MIR spectra, EMD works effectively because the spectra are dominated by the features from thermal emission of mineral grains and relatively free of contamination from grain reflection spectra or stellar photosphere fluxes. EMD is most effective at examining the spectral peak wavelength shifts due to changes in physical conditions (such as temperature) and stoichiometry such as Fe/Mg ratio for silicates. For NIR wavelength, the competing effect of scattering due to grain geometry and their thermal emission would require a distance measure that is scale invariant, such as Mahalanobis distance to analyze the stellar photosphere flux contribution or time-invariant distance measure such as Dynamic Time Warping (DTW) to account for the rapidly varying sky thermal and transmission background \citep{ackermann2010clustering,pandit2011comparative}. These additional distance measures can be easily added to the initial distance calculations (in the \sequencer step) and the rest of the \texttt{CLUES} analysis would remain the same. Thus, \texttt{CLUES} workflow can be easily adapted to other datasets with some modification to the pre-processing stage of the data processing pipeline and distance measures depending on the science application.

\subsection{Application to other datasets - Remote sensing multi/hyperspectral data}
In addition to astrophysical datasets, multispectral and hyperspectral (Visible to Near IR typically) datasets are becoming increasingly common for terrestrial remote sensing (e.g, NASA Plankton, Aerosol, Cloud, Ocean Ecosystem mission; Earth Surface Mineral Dust Source Investigation mission; \citealp{gorman2019nasa, green2020earth, guha2020mineral, qian2021hyperspectral}). These satellites are collecting large hypercubes (akin to \textit{JWST} NIR IFU datasets). Thus, this data can be readily analyzed with the \texttt{CLUES} workflow to infer distinct classes of surface mineralogy as well as land-water-cloud separation in an unsupervised approach. Our workflow would be directly complimentary to other recent approaches for unsupervised classification of remote sensing data using dimensionality reduction methods such as Principal Component Analysis and Uniform Manifold Approximation and Projection (UMAP) as well as supervised learning methods \citep{sousa2023topological,d2022automated,plebani2022machine}. Similarly, the \texttt{CLUES} workflow could be useful for analysis of planetary surface hyperspectral data e.g., MARS CRISM \citep{seelos2023crism, kumari2023mineral}, Moon Mineralogy Mapper \citep{kramer2011newer,gilmore2011superpixel}, and MESSENGER Mercury datasets \citep{murchie2015orbital}. Analogous to debris disk analysis, the \texttt{CLUES} workflow can serve as a first step to find exemplar spectra for detailed subsequent modeling using spectral libraries (e.g., USGS Tetracorder system \citealp{clark2003imaging}). While methods like UMAP has shown promise in analyzing some astrophysical spectral datasets  , our preliminary work on using UMAP for mineralogy classification in visible-Near IR hyperspectral satellite data did not show promising results. Thus, when considering complex, narrow band spectral features, a CLUES type approach may be more robust than time series/spectral mode decomposition approaches (e.g., UMAP, PCA).  

\subsection{CLUES vs just direct spectral fitting of the disks}
The main objective of \texttt{CLUES} is two-fold: (1) to use an unsupervised, data-driven approach to find exemplar systems that represent a border class and (2) to narrow down the vast parameter space of likely compositions for spectral modeling because narrowing down the parameter space amongst hundreds of compositions is computationally challenging for traditional parametric methods. 
More concretely, for a given disk spectrum, we can mix it in an emissivity library and only need to run \texttt{CLUES} once to obtain the original clustering of composition in the emissivity space. The emissivity library can be as large (with 1000s of components) as one would like. We can easily include pure materials with variations of porosity and grain size/shape distribution as well as mixed materials in various volume ratios. The scalability of \texttt{CLUES} allows us to quickly probe the degeneracy in compositions and grain size properties as a first step towards detailed spectral analysis and as an input to regression based approaches. Regression algorithms such as MCMC become numerically very expensive and unstable for very large spectral libraries with 100s to 1000s of components.

In addition, we can perform multiple levels of analyses with \texttt{CLUES}. For instance, we can consider a Spitzer IRS debris disk spectrum. Debris disk MIR spectra usually have two main components: the characteristic emission bands (often have FWHM of a tenth to a few microns) generated by small grains, and the broad, blackbody emission features (often have slowly varying slopes over 10-20 microns) generated from the ensemble of larger bodies. \texttt{CLUES} can be performed first on the raw MIR disk spectra to identify systems that possess a large amount of warm dust but are devoid of cold dust, from their counterparts. We could acquire demographic information on the overall dust temperature from this first level of analysis. After modeling the underlying black body emission and subtracting them from the original spectra to obtain residual spectra (see Figure~\ref{fig:continuum} for an example), we can run \texttt{CLUES} again on the specific MIR dust features to obtain more detailed information on grain species and properties. This type of analysis can be repeated again after removing the dominant dust component from the spectra to subsequently find 2nd order dust mineralogical components and so forth. The original \sequencer analysis has shown that such an approach can be feasible and is a very powerful approach to find small features \citep{Baron+21}. The exact utility and optimization of this procedure is beyond the scope of this study but will be described in detail in paper II. 

In the paper II, we will utilize the \texttt{CLUES} workflow to analyze the most comprehensive infrared, spectral debris disk catalog to-date, the \textit{Spitzer} IRS Debris Disk Catalog \citep{Chen+14, Mittal+15} with more than 500 disks, to conduct a systematic analysis to (1) identify the crust-like and mantle-like dust contents in the extrasolar debris disks, and to (2) investigate their relative dust fractions as a function of stellar properties such as stellar age, mass, and stellar luminosity. We will use the approach outline above - start with a global analysis of the disk structures (without continuum subtraction) followed by a continuum subtracted disk analysis and a single disk-by-disk mineralogy analysis akin to the result presented in this study. Furthermore, there are some Spitzer IRS specific challenges for the analysis with regards to normalization of 10, 20, and 30$\mu$m spectral features since they can represent multiple dust populations each with different dust mineralogy (see discussion in Section~\ref{sub:norm}). 

\section{Summary}\label{section:conclusion}
In this paper, we propose a novel, non-parametric classification workflow for spectral data -- \texttt{CLUES}. We demonstrate that such a workflow is widely applicable to NIR and MIR spectra in the field of planet formation and geological studies. \texttt{CLUES} is efficient in reducing dimensionality of spectra and narrowing down compositional parameter space that serves as good initial guesses to individual vetting of each spectrum. \texttt{CLUES} is also scalable to large data volume and deep data cubes as opposed to traditional parametric, component-based models. Our new workflow and results help set up a data-driven demographic analysis of the debris disk population in Paper II. We also posit that \texttt{CLUES} will be a widely useful tool in the era of \textit{JWST} where a substantial increase in MIR IFU data is anticipated. 

\section*{acknowledgements}
We thank Wen-Han Zhou and Winston Wu for helpful discussions.
CL and KW acknowledge support from STScI Director's Research Fund (DRF) and NASA FINESST grant No. 80NSSC21K1844 and 80NSSC22K1752 issued through the Mission Directorate. This research has received funding from the European Union's Horizon Europe research and innovation programme under the Marie Sk\l odowska-Curie grant agreement No.~101103114.

\software{\texttt{NumPy} \citep{numpy}, \texttt{Scipy} \citep{SciPy}, \texttt{Sequencer} \citep{Baron+21} }

\clearpage
\bibliographystyle{aasjournal}
\bibliography{main}
\end{document}